\documentclass[12pt,twoside]{article}
\usepackage{cmp209,amssymb,amsbsy,epsf}
\newcommand{\qref}[1]{(\ref{#1})}
\newcommand{\mshalf}{\hspace{-0.2mm}}
\newcommand{\msone}{\hspace{-0.4mm}}
\newcommand{\mstwo}{\hspace{-0.6mm}}
\newcommand{\psone}{\hspace{0.4mm}}
\newcommand{\fr}[2]{{\textstyle \frac{#1}{#2}}}
\newcommand{\prt}[2]{\frac{\partial \; #1}{\partial \; #2}}
\newcommand{\poch}[2]{\frac{{\rm d} \; #1}{{\rm d} \; #2}}

\newcommand{\sympm}[1]{ \mbox{\boldmath $ {\scriptstyle #1} $} }
\newcommand{\wh}[1]{\widehat{#1}}
\newcommand{\mframe}[1]{{#1}}
\newcommand{\ddd}{\! \cdot \! \cdot \! \cdot \!}
\renewcommand{\ae}{\varkappa}
\newcommand{\kt}{\tilde{\ae}}
\newcommand{\ktt}{\tilde{\tilde{\ae}}}
\newcommand{\barbars}{\bar{\! \bar{S}} {}}
\newcommand{\ttS}{\tilde{\! \tilde{S}} {}}
\newcommand{\ktx}{\tilde{\ae}^{\hspace{0.2mm}x}}
\newcommand{\kttx}{\tilde{\tilde{\ae}}{}^{\hspace{0.2mm}x}}
\newcommand{\kta}{\tilde{\ae}^{\hspace{0.2mm}a}}
\newcommand{\ktta}{\tilde{\tilde{\ae}}{}^{\hspace{0.2mm}a}}
\newcommand{\ktz}{\tilde{\ae}^{\hspace{0.2mm}z}}
\newcommand{\kttz}{\tilde{\tilde{\ae}}{}^{\hspace{0.2mm}z}}
\newcommand{\kora}{{}^k \mshalf}
\newcommand{\korb}{{}^k \msone}
\newcommand{\korc}{{}^k \mstwo}
\newcommand{\ham}{{\cal H}}
\newcommand{\krho}{\korc \rho}
\newcommand{\Fkor}{\korc {\cal F}}
\newcommand{\Ff}{{\cal F}}
\newcommand{\klang}{\korb \langle}
\newcommand{\korang}[1]{\klang #1 \rangle}
\newcommand{\kang}[1]{{\klang #1\rangle}}
\newcommand{\fiz}{\varphi^z}
\newcommand{\fix}{\varphi^x}
\newcommand{\fitn}[3]{ {}^{#1} \msone \varphi_{#2}^{#3} }
\newcommand{\fii}{\varphi}
\newcommand{\eps}{\varepsilon}
\newcommand{\ang}[1]{{\langle #1 \rangle}}
\newcommand{\ex}[1]{{\rm e}^{#1}}
\newcommand{\sh}{{\rm sh}}
\newcommand{\ch}{{\rm ch}}
\renewcommand{\th}{{\rm th}}
\newcommand{\Sp}{{\rm Sp}}
\newcommand{\etp}[1]{\big(E_{12}\big)_{\! #1}}
\newcommand{\etpn}[1]{(E_{1\mshalf 2})_{{}_{#1}}}
\newcommand{\omn}{\omega_n}
\newcommand{\vq}{\vec{q}}
\newcommand{\qomn}{\vq , \omn}
\newcommand{\ttau}{T_\tau \hspace{0.4mm} }
\newcommand{\ttaun}{T_\tau \hspace{0.4mm} }
\newcommand{\si}{\psone \sigma}
\newcommand{\sit}{\psone \tilde{\sigma}}
\newcommand{\attwo}[2]{ {}_{ \Big|
                   {}^{\scriptstyle { #1}}_{\scriptstyle { #2}}  } }

\newcommand{\RekPsi}{{\rm Re} \;\! \Psi}
\newcommand{\RekPsifull}{\RekPsi ^{zz} (0,E)}

\newcommand{\sdc}[1]{\langle #1 \rangle_{\msone \rho_{12}}^c }
\newcommand{\sdcom}[1]{ \big( \sdc{#1} \big)_{ \omn} }

\newcommand{\so}[1]{\langle #1 \rangle_{\msone \rho_{1}} }
\newcommand{\soc}[1]{\langle #1 \rangle_{\msone \rho_{1}}^c }
\newcommand{\socom}[1]{ \big( \soc{#1} \big)_{ \omn} }

\newcommand{\kacr}[2]{{\korang{ #1}}^{\!{}_{\scriptstyle c}}_{ #2}}
\newcommand{\acr}[2]{\langle { #1} {\rangle}
                    ^{{}_{\scriptstyle c}}_{ #2}}
\newcommand{\kar}[2]{{\korang{ #1}}_{ #2}}
\newcommand{\fgf}[5]{\Ff_{#1}^{(\mshalf #2 \mshalf)} ({}_{#3}^{#4,#5}) }
\newcommand{\fgfo}[3]{\fgf{1}{2}{#1}{#2}{#3} }
\newcommand{\fgfdr}[3]{\fgf{12}{1,1}{#1}{#2}{#3} }
\newcommand{\fgfdo}[3]{\fgf{12}{2,0}{#1}{#2}{#3} }
\newcommand{\fgfoom}[2]{\fgfo{\omn}{#1}{#2} }
\newcommand{\fgfdrom}[2]{\fgfdr{\omn}{#1}{#2} }
\newcommand{\fgfdoom}[2]{\fgfdo{\omn}{#1}{#2} }

\newcommand{\fgfm}[3]{{\wh \Ff}_{#1}^{(\mshalf #2 \mshalf)} (#3) }
\newcommand{\fg}{{\cal G}}
\newcommand{\fgm}{\wh{\fg}}

\newcommand{\fgmk}{\korb \fgm}
\newcommand{\bb}{ \!\backslash\;\! }
\newcommand{\od}{{\hspace{0.4mm}\prime}}

\newcommand{\tildL}{L}
\newcommand{\ii}{{\scriptstyle i}}
\newcommand{\fmi}{f}
\newcommand{\be}{\begin{equation}}
\newcommand{\ee}{\end{equation}}
\newcommand{\bea}{\begin{eqnarray}}
\newcommand{\eea}{\end{eqnarray}}

\newcommand{\brr}{ , }
\newcommand{\breps}{\{\eps\}}
\newcommand{\brepsn}{\{\msone\eps_\nu\msone\}}
\newcommand{\brepsr}{\{\msone\eps_r\msone\}}
\newcommand{\brepsnr}{\{\msone\eps_\nu\brr\eps_r\msone\}}

\newcommand{\hame}{{\rm H}(\!\breps\!)}
\newcommand{\hamen}{{\rm H}_{\nu}(\!\brepsn\!)}
\newcommand{\hamer}{{\rm H}_{r}(\!\brepsr\!)}
\newcommand{\epszero}{=\eps_r^a}
\newcommand{\kaptwo}{\;,\;\;{}^\nu\!\kt^a_{r,\tau}={}^\nu\!\kt^a_r}
\newcommand{\epsnbr}{(\!\{\msone\eps_\nu\msone\}\!)}
\newcommand{\epsnrbr}{(\!\{\msone\eps_\nu\brr\eps_r\msone\}\!)}
\newcommand{\etax}{\eta}

\title[Reference approach in theory of pseudospin systems]
{Reference approach in theory of pseudospin systems}
\author[R.R. Levitskii, S.I. Sorokov, O.R. Baran]
       {R.R. Levitskii, S.I. Sorokov, O.R. Baran}
\address{Institute for Condensed Matter Physics of the National Academy
of Sciences of Ukraine, 1 Svientsitskii Str., 79011 Lviv, Ukraine}
\begin{document}

\maketitle

\begin{abstract}
For theoretical description of pseudospin systems with essential
short-range and long-range interactions we use the method based on
calculations of the free energy functional  with taking into account
the short-range interactions within the reference approach in
cluster approximation. We propose a consistent formulation of the
cluster expansion method for quantum pseudospin systems. We develop a
method allowing one to obtain within the cluster approximation an
Ornstein-Zernike type equation for reference cumulant Green function
of an arbitrary order. In the two-particle cluster approximation we derived
an explicit expression for pair temperature cumulant Green function of the
reference system.
In the cluster random phase approximation we calculated and
studied thermodynamic characteristics, elementary excitation spectrum,
and integral intensities of the Ising model in transverse field.

\pacs 03.65.-w, 05.30.-d.
\end{abstract}

\section{Introduction}

Modern statistical theory of condensed media pays a great attention
to studies of ferroelectric and magnetic materials, described
by pseudospin models with essential short-range and long-range
interactions, especially of hydrogen-bonded ferroelectrics
\cite{r1,r2,r3,r4,r5,r6,r7,r8,r9,r10,r11,r12}
and low-dimensional magnets \cite{r13,pr4}. For
an adequate description of these objects, such an approach is required
that would allow to use different techniques for taking into account
short-range and long-range interaction. This is a typical mathematical
problem in theories of multiparticle systems. It has been successfully
solved in studies of equilibrium properties of classical systems
\cite{r15,r16,r17,r18,r19,r20,r21} and
metals \cite{r21,r22,r23,r26,r27}
with making use of the proposed in Refs.
\cite{r15,r16,r17,r18,r22,r26,r27}
approach. Within this approach, the long-range and short-range
interactions are  described in phase spaces of collective variables
and individual coordinates, respectively. The system with short-range
interaction is called then the reference system.

Using the idea of separating reference system
\cite{r15,r16,r17,r18}, in Refs. \cite{pr21,r30,r31,r32,r33}
a method  was proposed for description of pseudospin systems
with essential short-range and long-range interactions. This method is
based on calculation of the free energy functional with taking into
account the short-range interactions within the reference approach. In
Refs. \cite{pr21,r30,r33}
expansions of the free energy functional
and functionals of the temperature
cumulant Green functions (CGF)
in the inverse long-range interaction radius were
studied. For the first time there
has been performed a total summation of the
reducible in blocks diagrams  in the free energy functional
 and of non-compact diagrams
in functionals of CGFs for quantum pseudospin models. Expressions for the
free energy and temperature CGFs of the considered systems were obtained.
There was shown how to obtain consistent approximations for their
thermodynamic and dynamic characteristics, using classification of the
approximations for free energy functional
according to loop diagrams.

It should be noted that obtained in Refs.
\cite{pr21,r30,r31,r32,r33} general
expressions for thermodynamic and dynamic characteristics of
pseudospin systems with short-range and long-range interactions contain
thermodynamic and correlation functions  of the reference system. Hence, to
solve a general problem one needs to solve a reference one, that is, to
calculate free energy and CGFs of the reference system. The maximal order
of the reference CGFs depends on the order of the approximation for
the long-range interactions. Depending on the reference Hamiltonian, the
reference problem can be solved exactly (see, for instance,
\cite{pr4,pr63,r35,r36,r37,Oleg,r39}) or
approximately, with taking into account peculiarities of the reference
system. The best  description of the reference system for a wide
class of pseudospin models can be obtained on the basis of the cluster
expansions method (see
\cite{r1,r2,r3,r4,r5,r6,r7,r8,r9,r10,r11,r12,r40,r41,r42}).
In some papers \cite{r43,r44,r45,r46} this
method was successfully used for studies of disordered magnetic and
ferroelectric materials. Unfortunately, the cluster method was
correctly developed only for Hamiltonians with commuting  unary
and interaction parts. Mostly it was used for calculations of thermodynamic
characteristics of pseudospin models. In Refs.
\cite{r47,r48,r49} a problem of calculation
of distribution functions for Ising models
within the cluster approach was  considered but not solved completely.
 Equations for pair correlation
functions of the reference system (Ornstein-Zernike type equation) in
Refs. \cite{r47,r48} were not derived consistently
but constructed artificially.  The problem of calculation of
quasimomentum-dependent pair correlation functions was also considered in
Ref. \cite{r49} within the cluster approach.
The obtained there results are valid in paraphase only.
Later,  a method
was proposed \cite{pris,r51}, which allows one to obtain Ornstein-Zernike
type equations for an arbitrary order correlation functions of Ising models.
These equations for pair and three-particle correlation
functions were derived and solved within the two-particle cluster
approximation (TPCA).
It was shown that the proposed in Refs. \cite{pris,r51} cluster
approach to calculation of correlation functions of the reference Ising
models yields known exact results
\cite{pr63,r35} for pair and three-particle
correlation functions of the one-dimensional Ising model.

In Refs. \cite{r51,r52},
using the four-particle cluster approximation, pair
$\vq$-dependent correlation functions of deuterons were calculated for
KD$_2$PO$_4$ type ferroelectrics and ND$_4$D$_2$PO$_4$ type
antiferroelectrics. Dynamics of hydrogen-bonded ferroelectrics with taking
into account tunnelling effects was considered in Refs.
\cite{he32,he33,he34}  within the
proposed in \cite{he32} original approach.
For the first time it has been shown, that
in the reference
approach, with the short-range interactions  and tunnelling taken
into account in cluster approximation, the dynamic properties of the studied
systems are to a great extent determined by  an effective tunnelling
parameter, renormalized by the short-range interactions. Later, this
peculiarity  of the dynamic properties of hydrogen-bonded compounds was
noticed also in Ref. \cite{r56}.
Unfortunately, expressions for dynamic (at $\vq=0$
and $E=0$) and static characteristics, calculated in
Refs. \cite{he32,he33,he34}, turned to be inconsistent.
That results from the fact, that dynamic characteristics
were obtained using the method of two-time temperature Green functions,
equations for which were decoupled
in the spirit of Tyablikov approximation.
On this, the intracluster Green functions of the reference system were
connected only via the long-range interactions, whereas the short-range
correlations were not taken into account. Thus, there had not been
developed a method, which would allow one to consistently describe
thermodynamic and  dynamic characteristics of reference quantum pseudospin
models.

In the present paper, for theoretical description of pseudospin systems with
essential short-range and long-range interactions we shall use the developed
in Refs. \cite{pr21,r30,r31,r32,r33}
self-consistent reference approach. In Section 2 we shall briefly
consider the main results obtained within this
approach. Then, a consistent
formulation of cluster expansion method for
reference quantum pseudospin systems will
be given for the first time. We shall propose a method, allowing to
obtain Ornstein-Zernike type equations for reference temperature cumulant
Green functions of an arbitrary order within a cluster approximation.
An Ornstein-Zernike type equation for the pair correlator will be derived
and solved in the two-particle cluster approximation. The last Section is
devoted to investigation  of the Ising model in transverse field (IMTF)
within the cluster random phase approximation (CRPA)
using the obtained earlier in this paper results.

\section{Theory of pseudospin system with short-range
interactions \protect\newline
taken into account in reference approach}

We consider pseudospin systems with
short-range and long-range interactions, described by the Hamiltonian
\bea
\label{2.1}
\ham(\{\Gamma\})=-\beta H =
\sum_{\nu=1}^N \sum_{a} \Gamma^a_\nu S^a_\nu+\fr12
\sum_{\nu,\delta} \sum_{a,b}
K^{ab} S^a_\nu S^b_{\nu+\delta}
+\fr12 \sum _{\nu,\mu} \sum _{a,b}
J^{ab}_{\nu\mu} S^a_\nu S^b_\mu \; .
\eea
Here $K^{ab}$ and $J_{\nu\mu}^{ab}$ are the
short-range and long-range parts of the pair interactions.
$S_\nu^a$ ($a=x,y,z$ or $+,-,z$) are components of a normalized
($S^z=-1,1$) spin ${\vec S}$.
$\ham(\{\Gamma\})$ means
$\ham(\{\Gamma\})= \ham
(\Gamma^{a_1}_1,...,\Gamma^{a_1}_N,\Gamma^{a_2}_1,...,
\Gamma^{a_2}_N,\Gamma^{a_3}_1,...,\Gamma^{a_3}_N,)$.
Hereafter, the argument $\{\Gamma\}$ will be frequently omitted.
The factor $\beta=1/(k_BT)$, occurring in $\Gamma$, $K$, and $J$
(in the presented above form of the Hamiltonian $H$), will be written
explicitly only in some of final expressions.

After an identity transformation of the operators
$S_\nu^a=\ang{S_\nu^a} + \Delta S_\nu^a$
in the last term of Hamiltonian \qref{2.1}, which describes the
long-range interactions between pseudospins, we obtain
\be
\label{2.3}
\ham(\{\Gamma\})=\kora {\cal H}(\{\ae\})
-\fr12 \sum_{\nu,\mu} \sum_{a,b}
J^{ab}_{\nu\mu} \ang{S^a_\nu} \ang{S^b_\mu}
+\fr12 \sum_{\nu,\mu} \sum_{a,b}
J^{ab}_{\nu\mu} \Delta S^a_\nu \cdot \Delta S^b_\mu \; .
\ee
The first term in \qref{2.3} describes the short-range interactions
between pseudospins placed in a field created by
the long-range interactions and by  $\Gamma_\nu^a$
\bea
\label{2.4}
&&
\kora \ham (\{\ae\})= -\beta \!\cdot\! \korb H (\{\ae\})=
\sum_{\nu=1}^N
\sum_{a} \ae^a_\nu S^a_\nu+\fr12 \sum_{\nu,\delta}
\sum_{a,b} K^{ab} S^a_\nu S^b_{\nu+\delta} \;\!;
\\ &&
\label{2.5}
\ae_\nu^a=\Gamma_\nu^a +\sum_{\mu=1}^N \sum_b
J_{\nu\mu}^{ab}\ang{S_\mu^b} \;\!.
\eea
The Hamiltonian $\korb H (\{\ae\})$ is called the reference Hamiltonian
\cite{pr21}. The argument $\{\ae\}$ will be dropped often.
Let us note that in the mean field approximation
(MFA) over the long-range interactions, the
last term in \qref{2.3} is neglected.

Our main task is to calculate the free energy
\bea
\label{2.6}
F \big(\{\Gamma\}\big)= -k_BT \ln Z \big(\{\Gamma\}\big) \;, \qquad
Z \big(\{\Gamma\}\big) = \Sp \; \ex{ \ham}
\eea
and pair temperature cumulant Green functions
\bea
\label{2.7}
b^{(2)} \big({}^{a_1}_{\nu_1,\tau_1}
\bb {}^{a_2}_{\nu_1,\tau_2} \bb \big)
= \acr{ \ttau \; \ttS_{\nu_1}^{a_1} (\tau_1) \;
\ttS_{\nu_2}^{a_2} (\tau_2) }{ \rho }
\eea
for the models, described by Hamiltonian \qref{2.1}. Here
\bea
\label{2.8}
\ttS{}_{\nu}^{a}(\tau)= {\rm e}^{-\tau \ham} \;\!
S_\nu^a \; {\rm e}^{\tau \ham} \;;
\eea
the averaging is performed with
the density matrix
\bea
\label{2.9}
\rho=
\rho(\{\Gamma\})=[{Z(\{\Gamma\})}]^{-1} \cdot {{\rm e}^{\ham}} \;.
\eea
For the sake of convenience, in our calculations we  use not the
free energy $F(\{\Gamma\})$ but the ${\cal F}(\{\Gamma\})$--function
(logarithm of the partition function).

According to the theory proposed in Ref.~\cite{pr21}, to solve
the formulated problem, one should calculate the $\Fkor$-function
\be
\label{2.18}
\Fkor(\{\ae\})=
\ln \korb Z (\{\ae\}) ; \qquad
\korb Z (\{\ae\}) =
{\rm Sp} \; {\rm e}^{\kora {\ham}}
\ee
and CGFs
\bea
\label{2.19}
&&
\kora b^{(l)}
\big({}^{a_1}_{\nu_1,\tau_1} \bb {}^{a_2}_{\nu_2,\tau_2}
\bb ... \bb {}^{a_l}_{\nu_l,\tau_l} \bb  \big) =
\kacr{  \ttau \tilde{S}_{\nu_1}^{a_1} (\tau_1)
\tilde{S}_{\nu_2}^{a_2} (\tau_2) ...
\tilde{S}_{\nu_l}^{a_l} (\tau_l) } { \krho } \;\!;
\\[7pt] && \hspace{-6mm}
\label{2.20}
\tilde{S}_\nu^{a} (\tau) =\ex{-\tau \cdot \kora \ham} \;
S_\nu^{a} \; \ex{\tau \cdot \kora \ham}, \qquad
\krho= \krho (\{\ae\}) =
[{\kora Z(\{\ae\}) }]^{-1} \cdot { \ex{ \kora \ham } }
\eea
of the reference system \qref{2.4}.
In the present paper the $\Fkor$-function and pair CGFs
will be calculated within the
two-particle cluster approximation.

Assuming that the reference problem is solved in Ref.~\cite{pr21}
the expansion of the free energy functional in the inverse radius of
the long-range interactions was studied, for the systems
described by the Hamiltonian \qref{2.1} and expressions
for temperature Green functions were obtained.
Here we present only some of their results for non-uniform
fields ($\Gamma_\nu^a=\Gamma^a$) up to $r_0^{-d}$ in the long-range
interactions.

The ${\cal F}$-function of the considered system reads
\bea
\label{2.28} \!\!\!\!\!
{\cal F}(\!\{\Gamma\}\!) \!=\! \Fkor(\!\{\ae\}\!)
-\fr N2 \sum_{a,b} J^{ab}_0 \ang{S^a}_{\! \rho}
\ang{S^b}_{\! \rho} -
\fr12 \! \sum_{\omega_n,\vq} \ln \det \!
\Big[\wh{1}- \kora \wh{b}^{(2)}(\vq, \omega_n)
\wh{J}(\vq)\Big],
\eea
where
\bea
\label{2.30}
J^{ab}_0=J^{ab}( \vq=0 ), \qquad
\ae^a=\Gamma^a + \sum_b J^{ab}_0 \ang{S^b}_{\! \rho} \; ;
\eea
$\kora \wh{b}^{(2)}(\vq, \omega_n)$
and $\wh{J}(\vq)$ are matrices $3\times3$ in the indices $a$, $b$;
their elements are Fourier transforms
$\kora b^{(2)}({}^{a\;b}_{\vq,\omn})$ and $J^{ab}(\vq)$
of the pair CGFs of the reference system $\kora b^{(2)}
\big({}^{a}_{\nu,\tau_1} \bb {}^{b}_{\mu,\tau_2} \bb \big)$
and of the long-range interactions $J_{\nu\mu}^{ab}$
(for uniform fields ($\ae^a_\nu=\ae^a$)).

For the pair CGFs
$b^{(2)}\big({}^{a}_{\nu,\tau_1} \bb {}^{b}_{\mu,\tau_2} \bb \big)$
in the frequency-momentum space, the
following relation \cite{pr21} holds
\be
\label{2.31}
\wh{b}^{(2)}(\vq,\omn)
=\Big[1-\wh M( \vq, \omn )\wh J(\vq) \Big]^{-1}
\wh M(\vq,\omn)
\ee
Here we use the notation
\bea
\label{2.32}
&& \hspace{-15mm}
M \big({}^{a \; b}_{\vq,\omn} \big) =
\kora b^{(2)}\big({}^{a\;b}_{\vq,\omn})
+ \fr 1{2N} \sum_{\{a,b\}} \sum_{\{\vq_i, \omega_{n_i} \}}
\kora b^{(3)} \big( {}^{a}_{\nu_0, \tau_0} \bb
{}^{a_1}_{\vq_1, \omega_{n_1}} \bb
{}^{a_2}_{\vq_2, \omega_{n_2}} \bb \big)
\\[5pt] \nonumber &&
\times
\kora b^{(3)} \big({}^{b}_{\nu_0, \tau_0} \bb
{}^{b_1}_{-\vq_1, -\omega_{n_1}} \bb
{}^{b_2}_{-\vq_2, -\omega_{n_2}} \bb \big)
{\cal R}\big({}^{a_1 \; b_1}_{-\vq_1, -\omega_{n_1}} \big)
{\cal R}\big({}^{a_2 \; b_2}_{-\vq_2, -\omega_{n_2}} \big)
\\[5pt] \nonumber &&
\times
\delta (\vq + \vq_1+\vq_2)
\delta (\omega_n+\omega_{n_1}+\omega_{n_2}) +
\\[5pt] \nonumber &&
+\fr1{2N} \sum_{a_1,a_2} \sum_{\vq_1, \omega_{n_1}}
\kora b^{(4)}
\big({}^{a}_{\nu_0,\tau_0} \bb {}^{b}_{-\vq, -\omega_n} \bb
{}^{a_1}_{\vq_1,\omega_{n_1}} \bb
{}^{a_2}_{-\vq_1,-\omega_{n_1}} \bb  \big)
{\cal R}\big({}^{a_1 \;a_2}_{-\vq_1, -\omega_{n_1}} \big) \;\!,
\eea
where
\bea
\label{2.33} \!\!
{\wh {\cal R}}(\vq, \omn )=
{\wh J}(\vq) \Big[1- \kora {\wh b}^{(2)}
(\vq, \omn ) {\wh J}(\vq)\Big]^{-1} \qquad\!\!\!
(\nu_0=0, \; \tau_0=0, \; \omn=2\pi n \beta^{-1})
\eea
is the Fourier transform
of the effective interaction in the considered system, whereas
$ \kora b^{(3)} \big( {}^{a}_{\nu_0,\tau_0} \bb
  {}^{a_1}_{\vq_1, \omega_{n_1}} \bb
  {}^{a_2}_{\vq_2, \omega_{n_2}} \bb \big) $
and
$ \kora b^{(4)} \big({}^{a}_{\nu_0,\tau_0} \bb
  {}^{b}_{-\vq, -\omega_n} \bb {}^{a_1}_{\vq_1,\omega_{n_1}} \bb
  {}^{a_2}_{-\vq_1,-\omega_{n_1}} \bb \big) $
are Fourier transforms of the three and four-particle CGFs of
the reference system, respectively.

We also present here an expression for the order parameter \cite{pr21}
\bea
\label{2.34}
\ang{S^a}=
\ang{S^a}_{\rho}=
\kang{S^a}_{\krho}
+\fr{1}{2N}\sum_{a_1,a_2}\sum_{\vq,\omn} \kora b^{(3)}
\big( {}^{a}_{\nu_0,\tau_0} \bb {}^{a_1}_{\vq,\omn} \bb
{}^{a_2}_{-\vq,-\omn} \bb \big)
{\cal R}\big({}^{a_1 \; a_2}_{-\vq ,-\omn} \big) \;\!.
\eea

Hence, we have general expressions for the free energy
($F=-k_BT{\cal F}$) and pair CGFs,  and the equation for the order
parameter in the  $r_0^{-d}$ approximation. These expressions contain the
free energy and correlation functions of the reference system.

\section{Two-particle cluster approximation for short-range
          interactions}
\setcounter{equation}{0}

\subsection{Problem formulation}

Our task in this section is to obtain the
$\Fkor(\!\{\ae\}\!)$-function, parameters
$\kang{S_\nu^a}_{\krho}$ and pair  cumulant Green
functions $\kora b^{(2)}
\big({}^{a_1}_{\nu,\tau_1} \bb {}^{a_2}_{\mu,\tau_2} \bb \big)$
of the reference pseudospin system, described by Hamiltonian
$\korb H(\{\ae\})$ \qref{2.4}, in the two-particle cluster
approximation in the short-range interactions.

Let us define the functional of the partition function logarithm
(the $\Fkor(\{\eps\})$-functional) of the reference model as
\cite{pr21}
\bea
\label{d9}
&&
\Fkor\big(\{\eps\},\{\ae\}\big) = \ln
\korc Z\big(\{\eps\},\{\ae\}\big) \; ;
\\[6pt] && \nonumber
\korc Z \big(\{\eps\},\{\ae\}\big) =
\Sp \Big[ \ex{\hame} \;\! \ttau \exp \Big(\int_0^1 {\rm d}\tau \;
\kora \ham\big(\tau,\{\ae\}\big) \Big) \Big] \;,
\eea
where
\bea
&&
\label{d11} \hspace{-16mm}
\kora \ham\big(\tau,\{\ae\}\big)=
\sum_a \sum_{\nu=1}^N \ae^a_{\nu,\tau} S^a_{\nu,\tau} + \fr 12
\sum_{a,b} \sum_{\nu,\delta}
K^{ab} S^a_{\nu,\tau} S^b_{\nu+\delta,\tau} \; ;
\\ &&
\label{d10} \hspace{-16mm}
\hame=\sum_{\nu=1}^N \hamen \;\! ; \qquad\!\!
\hamen=\sum_a \eps_\nu^a \;\! S_\nu^a \;\! ; \qquad\!\!
A_\tau= \ex{-\tau \hame} \; A \;  \ex{\tau \hame} \;\! .
\eea
Dependence $\ae^a_{\nu,\tau}$ on $\tau$ here is necessary to perform
functional differentiation with respect to $\ae^a_{\nu,\tau}$
\cite{pr21}.
It should be noted, that since spin operators at different sites commute,
and $\ae^a_\nu$ is a scalar, the quantity $A_{\tau}$
(if $A=S^a_\nu$ or $\ae^a_\nu$) can be written as

\vspace*{2mm}
\[
A_{\tau} = \ex{-\tau \hamen} \; A \;  \ex{\tau \hamen} \; .
\]

Starting from \qref{d9}, we introduce functionals of CGFs of the
reference system
\bea
\label{d12a}
&&
\kora b^{(l)}
\big({}^{a_1}_{\nu_1,\tau_1} \bb {}^{a_2}_{\nu_2,\tau_2}
\bb...\bb {}^{a_l}_{\nu_l,\tau_l} \bb \{\eps\} \big) =
\kacr{ \ttau S_{\nu_1,\tau_1}^{a_1}  S_{\nu_2,\tau_2}^{a_2}
...  S_{\nu_l,\tau_l}^{a_l} } { \krho(\!\{\eps\}\!) } \;\!;
\\[7pt] && \nonumber
\krho(\!\{\eps\}\!)= \frac{1}{\korc Z(\{\eps\}, \{ \kt \})}
\; \ex{\hame}\;\!  \exp \Big[ \int_0^1 {\rm d} \tau \;
\kora \ham \big(\tau, \{ \ae \}\big) \Big]  \; .
\eea
They will be found using
\bea
\label{d13}
\kora b^{(l)}
\big({}^{a_1}_{\nu_1,\tau_1} \bb {}^{a_2}_{\nu_2,\tau_2}
\bb...\bb {}^{a_l}_{\nu_l,\tau_l} \bb \{\eps\} \big) =
\frac{\delta}{\delta \ae_{\nu_1,\tau_1}^{a_1} }
\frac{\delta}{\delta \ae_{\nu_2,\tau_2}^{a_2} }
\ldots
\frac{\delta}{\delta \ae_{\nu_l,\tau_l}^{a_l} }
\;\! \Fkor\big(\{\eps\},\{\ae\}\big) \; .
\eea

According to \cite{pr21}, the following relations between the
$\Fkor(\{\ae\})$-function \qref{2.18}
and temperature CGFs \qref{2.19} and their functionals hold
\bea
&&
\label{d7}
\Fkor\big(\{\ae\}\big)= \Fkor\big(\{\eps\},\{\ae\}\big)
\attwo{\ae^a_{\nu,\tau}=\ae^a_\nu }{\eps_\nu^a=0} \; ;
\\[9pt] &&
\label{d8}
\kora b^{(l)}
\big({}^{a_1}_{\nu_1,\tau_1} \bb {}^{a_2}_{\nu_2,\tau_2}
\bb...\bb {}^{a_l}_{\nu_l,\tau_l} \bb \big) =
\kora b^{(l)}
\big({}^{a_1}_{\nu_1,\tau_1} \bb {}^{a_2}_{\nu_2,\tau_2}
\bb...\bb {}^{a_l}_{\nu_l,\tau_l} \bb \{\eps\} \big)
\attwo{\ae^a_{\nu,\tau}=\ae^a_\nu }{\eps_\nu^a=0} \; .
\eea

\noindent
That is, calculation of the $\Fkor(\{\ae\})$-function and
temperature CGFs is reduced to calculation of the
$\Fkor(\{\eps\},\{\ae\})$-functional.

\subsection{Cluster approximation. Free energy}

Let us calculate now the
$\Fkor\big(\{\eps\},\{\ae\}\big)$-functional in the two-particle cluster
approximation. We perform a cluster expansion, with the lattice
being divided into the two-particle clusters \cite{pris,prhe,mpri}.
As
$
\sum_a \fitn{r}{\nu,\tau}{a} S_{\nu,\tau}^a
$
we denote an operator of the effective field created by the site
$r$ and acting on the site $\nu$, provided that the site $r$ is a
nearest neighbour of the site $\nu$ $(r \in \pi_\nu)$.
Obviously, the number of fields acting on an arbitrary site $\nu$
\[
\sum_{r \in {\textstyle \pi_\nu}}
\sum_a \fitn{r}{\nu,\tau}{a} S_{\nu,\tau}^a
\]
is $z$ ($z$ is the nearest neighbours number).
After an identity transformation, the reference Hamiltonian
\qref{d11} takes the form
\bea
\label{d15}
\kora \ham\big(\tau,\{\ae\brr\fii\} \big) =
\sum_{\nu}\ham_\nu \big(\tau,\{ \kt_\nu \}\big) +
\sum_{(\nu,r)}
{\cal U}_{\nu r}
\big(\tau,\{\fitn{r}{\nu}{}\brr\fitn{\nu}{r}{}\} \big) \; ,
\eea
where
\bea
\label{d16}
&& \msone
\ham_\nu \big(\tau,\{\kt_{\nu}\}\big) =
\sum_{a} \kt_{\nu,\tau}^a S_{\nu,\tau}^a \;; \qquad
\kt_{\nu,\tau}^a = \ae_{\nu,\tau}^a
+ \sum_{r \in {\textstyle \pi_\nu}} \fitn{r}{\nu,\tau}{a} \;;
\\
\label{d17}
&& \msone
{\cal U}_{\nu r}
\big( \tau,\{\fitn{r}{\nu}{}\brr\fitn{\nu}{r}{}\} \big)=
\sum_{a} \Big( - \fitn{r}{\nu,\tau}{a} S_{\nu,\tau}^a
- \fitn{\nu}{r,\tau}{a} S_{r,\tau}^a + \sum_{b}
K^{ab} S_{\nu,\tau}^a S_{r,\tau}^b    \Big) \;. \qquad
\eea
$\ham_\nu \big(\tau, \{\kt_\nu\}\big)$ means $\ham_\nu = \ham_\nu
(\kt_{\nu,\tau}^{a_1},\kt_{\nu,\tau}^{a_2},\kt_{\nu,\tau}^{a_3})$.
Hereafter, the arguments $\{\kt_\nu\}$,
$\{\fitn{r}{\nu}{}\}$ will frequently omitted.

Let us present the $\Fkor(\{\eps\})$-functional
\qref{d9} as
\bea
\label{d18}
&& \hspace{-15mm}
\Fkor \big(\{\eps\},\{ \ae\brr \fii\}\big)
\\ && \nonumber \hspace{-10mm}
= \ln \Sp \Big\{ \ex{\hame} \;\! \ttau
\exp\Big[\sum_{\nu=1}^N \int_0^1 {\rm d}\tau \; \ham_\nu
\big(\tau\big) \Big]
\exp\Big[\sum_{(\nu,r)} \int_0^1 {\rm d}\tau \;
{\cal U}_{\nu r} \big(\tau\big)  \Big]   \Big\}
\\ && \nonumber \hspace{-10mm}
= \sum_\nu \Ff_\nu \big(\brepsn, \{ \kt_\nu \}\big)
+ \ln  \langle
\ttaun \exp \Big(
\sum_{(\nu,r)} \int_0^1 {\rm d}\tau \; {\cal U}_{\nu r} \big(\tau\big)
\Big)
\rangle _{\rho_0(\!\{\eps\}\!)} \; .
\eea
Here $\Ff_\nu(\brepsn)$ is the so-called single-particle
intracluster $\Ff(\{\eps\})$-functional
\bea
\label{d20}
&&
\Ff_\nu \big(\brepsn , \{ \kt_\nu \}\big) =
\ln Z_\nu \big(\brepsn , \{ \kt_\nu \}\big) \;\; ;
\\[7pt]
\label{d21}
&&
Z_\nu \big(\brepsn , \{ \kt_\nu \} \big) =
\Sp_{\sympm{S}_\nu}  \Big\{ \ex{\hamen} \;\! \ttau
\exp \Big[ \int_0^1 {\rm d} \tau \;
\ham_\nu \big(\tau, \{ \kt_\nu \}\big) \Big] \Big\} \;
\eea
and averaging is performed with the functional of the density matrix
\bea
\label{d19} && \hspace{-10mm}
\rho_0(\!\breps\!) \!=\! \prod_\nu \rho_\nu (\!\brepsn\!) \;\!;
\\ && \label{d19a} \hspace{-10mm}
\rho_{\nu}(\!\brepsn\!)= \frac{1}{Z_\nu(\brepsn, \{ \kt_\nu \})}
\; \ex{\hamen} \;\! \exp \Big[ \int_0^1 {\rm d} \tau \;
\ham_\nu \big(\tau, \{ \kt_\nu \}\big) \Big]  \;\! .
\eea

We restrict our consideration by the first order of the cluster
expansion \cite{pris,prhe,mpri};
this corresponds to the two-particle cluster
approximation. Then the $\Fkor(\!\{\eps\}\!)$-functional becomes a
sum of the single- and two-particle intracluster
$\Ff(\!\{\eps\}\!)$-fun\-ctionals
\bea
\label{d22}
\Fkor\big(\{\eps\}, \{ \ae\brr\fii \}\big) =\!
\sum_\nu \Ff_\nu \big(\brepsn ,\{ \kt_\nu \}\big) +
\sum_{(\nu,r)} \ln  \langle \ttaun \msone
\exp \! \Big(\int_0^1 \! {\rm d}\tau \;
{\cal U}_{\nu r} \big(\tau\big) \Big)
\rangle _{\rho_0(\!\{\eps\}\!)}  \quad
\\ \nonumber \hspace{6mm}
= (1-z) \sum_\nu \Ff_\nu \big(\brepsn, \{ \kt_\nu \} \big)+
\fr{1}{2} \sum_{\nu ,r}
\Ff_{\nu r}\big(\brepsnr ,
\{ {}^r \kt_\nu\brr {}^\nu \kt_r \} \big) \; .
\eea
The two-particle $\Ff_{\nu r}(\brepsnr)$-functional reads
\bea
\label{d23}
&& \hspace{-15mm}
\Ff_{\nu r}
\big(\brepsnr,
\{ {}^r \! \kt_\nu \brr {}^\nu \! \kt_r \}\big) \!=
\ln Z_{\nu r}
\big(\brepsnr, \{ {}^r \! \kt_\nu \brr {}^\nu \! \kt_r \} \big) ;
\\[8pt] && \hspace{-15mm}
\label{d24}
Z_{\nu r}
\big(\brepsnr
\big) \!= \msone
\Sp_{\sympm{S}_1,\sympm{S}_2} \msone
 \Big\{ \ex{\hamen+\hamer} \!\; \ttau
\exp \! \Big[ \! \int_0^1 \! {\rm d} \tau \;\msone
\ham_{\nu r}
\big(\tau, \{ {}^r \! \kt_\nu\brr {}^\nu \! \kt_r \} \msone \big)
\Big] \! \Big\} ;
\\[8pt] && \hspace{-15mm}
\label{d25}
\ham_{\nu r}
\big(\tau, \{ {}^r \! \kt_\nu\brr {}^\nu \! \kt_r \}\big) =
\ham_{\nu}\big(\tau,  \{ \kt_\nu \}\big) +
\ham_{r}\big(\tau, \{ \kt_r \}\big) +
{\cal U}_{\nu r}\big(\tau\big)
\\[5pt] && \nonumber
= \sum_{a} \Big[
{}^r \! \kt^a_{\nu,\tau} S^a_{\nu,\tau} + {}^\nu \! \kt^a_{r,\tau}
S^a_{r,\tau} +
\sum_b K^{ab} S^a_{\nu,\tau}  S^b_{r,\tau} \Big] \;;
\\[5pt] && \hspace{-15mm}
\label{d26}
{}^r \! \kt^a_{\nu,\tau} = \kt^a_{\nu,\tau} -
{}^r \! \fii^a_{\nu,\tau} =
\ae^a_{\nu ,\tau} +
\sum_{ {}
^{\scriptstyle r' \in {\textstyle\pi_\nu}}
_{\scriptstyle r' \neq  \;\! r} }
\fitn{r'\!}{\nu,\tau}{a} \;.
\eea

Putting $\eps^a_\nu=0$ ($\ae^a_{\nu,\tau}=\ae^a_{\nu}$,
$\fitn{r}{\nu,\tau}{a}=\fitn{r}{\nu}{a}$, see also \qref{d7}),
and going to the uniform fields case $\ae^a_{\nu}=\ae^a$
($\fitn{r}{\nu}{a}=\fii^a$), from  \qref{d22}
we obtain the $\Fkor$-function of the reference system in the TPCA
for the uniform fields
\bea
\label{d22a} && \hspace{-10mm}
\Fkor\big(\{ \ae \brr \fii \}\big) =
(1-z) N \Ff_1 \big(\{ \kt \} \big)+
\fr{Nz}{2}
\Ff_{12}\big(\{ \ktt \} \big) \; ;
\\[8pt] && \hspace{-10mm}
\label{d20a}
\Ff_1\big(\{ \kt \}\big) =
\ln Z_1\big(\{ \kt \}\big) \; ; \qquad
Z_1\big(\{ \kt \} \big) =
\Sp_{\sympm{S}_1} \ex{\ham_1(\{ \kt \})} \;;
\\[5pt] && \hspace{-10mm}
\label{d16a}
\ham_1 \big( \{\kt \}\big) =
\sum_{a} \kt^a S_{1}^a \;\; ; \qquad
\kt^a = \ae^a + z \fii^a \;;
\\[5pt]
\label{d23a}
&& \hspace{-10mm}
\Ff_{12}\big(\{ \ktt \}\big) =
\ln Z_{12} \big(\{ \ktt \}\big) \; ; \qquad
Z_{12} \big(\{ \ktt \} \big) =
\Sp_{\sympm{S}_1, \sympm{S}_2} \ex{\ham_{12}(\{ \ktt \})} \;;
\\[5pt] && \hspace{-10mm}
\label{d25a}
\ham_{12} \big( \{\ktt \}\big) =
\sum_{a} \big[ \ktt{}^a \big( S_{1}^a+S_{2}^a \big) +
\sum_b K^{ab}S_{1}^a S_{2}^b \big]
\; ; \qquad
\ktt{}^a = \ae^a + (z-1) \fii^a . \qquad
\eea

\subsection{System of equations for single-particle distribution
functions and variational parameters}

Let us find now equations  for functionals
$ \korang{\ttau S^a_{\nu,\tau} }_{\krho(\!\{\eps\}\!)} $
and for cluster fields $ {}^r \fii^a_{\nu ,\tau}$. From
\qref{d13} we obtain

\bea
\label{d27}
\kar{\ttau S_{\mu,\tau}^{a} }{\krho(\!\{\eps\}\!)}
= \frac{ \partial \; \Fkor(\!\{\eps\}\!) }
{ \partial \; \ae_{\mu,\tau}^{a} } +
\sum_{\nu} \sum_{r \in {\textstyle \pi_\nu}} \sum_{b}
\int_0^1 {\rm d} \tau' \;
\frac{ \partial \; \Fkor(\!\{\eps\}\!) }
{ \partial \; {}^r \fii_{\nu,\tau'}^{b} } \cdot
\frac{ \delta \;  \fitn{r}{\nu,\tau'}{b} }
{ \delta \; \ae_{\mu,\tau}^{a} } \; .
\eea
Starting from the fact that the $\Fkor(\!\{\eps\}\!)$-functional
\qref{d22} is a sum of the single- and two-particle
$\Ff(\!\{\eps\}\!)$-functionals,
we get
\bea
\label{d28}
&& \hspace{-17mm}
\frac{ \partial \; \Fkor\big(\!\{\eps\}\!\big) }
{ \partial \; \ae^a_{\nu,\tau} } =
(1\!-\!z) \;\! \Ff_\nu^{(1)} \big(  \;{}^a_\tau  \big|
\brepsn \! \big) +
\sum_{r \in {\textstyle \pi_\nu}}
\Ff_{\nu r}^{(1,0)} \big( \; {}^a_\tau
\big| \brepsnr \! \big) \;;
\\
\label{d29}
&& \hspace{-17mm}
\frac{ \partial \; \Fkor\big(\!\{\eps\}\!\big) }
{ \partial \; {}^{r_1} \fii^a_{\nu,\tau} }
\!=\! (1\!-\!z) \!\; \Ff_\nu^{(1)}
\big( \; {}^a_\tau  \big| \brepsn \! \big)
\!+\!\sum_{ {}
^{\scriptstyle r \in {\textstyle\pi_\nu}}
_{\scriptstyle r \neq  \;\! r_1 } }
\Ff_{\nu r}^{(1,0)} \big( \; {}^a_\tau  \big| \brepsnr \!\big)
, \;\;\;\;\;\;\;
(r_1 \!\in\! {\textstyle\pi_\nu}) .
\eea
Here we introduce the notations
\bea
\label{d30}
&& \hspace{-15mm}
\Ff_\nu^{(k)} \big(\; {}^{ a_1 }_{ \tau_1 } \!\big| \;\;
{}^{ a_2 }_{ \tau_2 } \!\big| \;\; \ddd
\;\; {}^{ a_k }_{ \tau_k } \! \big| \;
\brepsn, \{ \kt_\nu \} \big) =
\frac{\partial}{\partial \; \kt_{\nu,\tau_1}^{a_1}} \cdot
\frac{\partial}{\partial \; \kt_{\nu,\tau_2}^{a_2}} \ddd
\frac{\partial}{\partial \; \kt_{\nu,\tau_k}^{a_k}} \;\!
\Ff_\nu\big(\brepsn, \{ \kt_\nu \} \big) \; , \quad
\\[7pt]
\label{d31}
&& \hspace{-15mm}
\Ff_{\nu r}^{(k,l)} \big( \;\;
{}^{ a_1 }_{ \tau_1 } \! \big| \;\;
{}^{ a_2 }_{ \tau_2 } \! \big| \;\;      \ddd \;\;
{}^{ a_k }_{ \tau_k } \! \big| \;\;
{}^{ a'_1 }_{ \tau'_1 } \! \big\| \;\;
{}^{ a'_2 }_{ \tau'_2 } \! \big\| \;\;    \ddd \;\;
{}^{ a'_l }_{ \tau'_l }  \big\| \;
\brepsnr, \{ {}^r \kt_\nu\brr {}^\nu \kt_r \} \big)
\\[5pt] && \nonumber  =
\frac{\partial}{\partial \; {}^r \mshalf \kt_{\nu,\tau_1}^{a_1} }
\ddd
\frac{\partial}{\partial \; {}^r \mshalf \kt_{\nu,\tau_k}^{a_k} }
\cdot
\frac{\partial}{\partial \; {}^\nu \mshalf \kt_{r,\tau'_1}^{a'_1} }
\ddd
\frac{\partial}{\partial \; {}^\nu \mshalf \kt_{r,\tau'_l}^{a'_l} }
\;\! \Ff_{\nu r} \big(\brepsnr,
\{ {}^r \mshalf \kt_\nu\brr {}^\nu \mshalf \kt_r \} \big) \;\! .
\eea
From the explicit form of the intracluster
$\Ff(\{\eps\})$-functionals \qref{d20}, \qref{d23} it follows that
\bea
\label{d36}
&& \hspace{-10mm}
\Ff_\nu^{(k)} \big(\; {}^{ a_1 }_{ \tau_1 } \!\big| \;\;
{}^{ a_2 }_{ \tau_2 } \!\big| \;\;       \ddd
\;\; {}^{ a_k }_{ \tau_k } \! \big| \; \brepsn, \{ \kt_\nu \} \big)
=
\acr{\ttau \; S^{a_1}_{\nu \tau_1} \; S^{a_2}_{\nu \tau_2} \;\ddd\;
S^{a_k}_{\nu \tau_k}  }   {\rho_\nu(\!\brepsn\!)} \; ;
\\ [9pt]
\label{d37}
&& \hspace{-10mm}
\Ff_{\nu r}^{(k,l)} \big( \;\;
{}^{ a_1 }_{ \tau_1 } \! \big| \;\;
{}^{ a_2 }_{ \tau_2 } \! \big| \;\;      \ddd \;\;
{}^{ a_k }_{ \tau_k } \! \big| \;\;
{}^{ a'_1 }_{ \tau'_1 } \! \big\| \;\;
{}^{ a'_2 }_{ \tau'_2 } \! \big\| \;\;    \ddd \;\;
{}^{ a'_l }_{ \tau'_l }  \big\| \;
\brepsnr, \{ {}^r \kt_\nu\brr {}^\nu \kt_r \} \big)
\\[5pt] && \nonumber =
\acr{\ttau \; S^{a_1}_{\nu \tau_1} \; S^{a_2}_{\nu \tau_2} \;\ddd\;
S^{a_k}_{\nu \tau_k} \; S^{a'_1}_{r \tau'_1} \;
S^{a'_2}_{r \tau'_2} \;\ddd\; S^{a'_l}_{r \tau'_l}}
{\rho_{\nu r}(\!\brepsnr\!)} \; ,
\eea
where the averagings are performed with the density matrix
functionals \qref{d19a} and
\bea
\label{d35}
\rho_{\nu r}(\!\brepsnr\!)= \frac
{\; \ex{\hamen+\hamer} }{Z_{\nu r}
(\brepsnr, \{ {}^r \mshalf \kt_\nu \brr {}^\nu \mshalf \kt_r \})}
\; \exp \! \Big[ \int_0^1 {\rm d} \tau \;
\ham_{\nu r} \big(\tau, \{ {}^r \mshalf \kt_\nu\brr
{}^\nu \mshalf \kt_r \}\big) \Big]
\!\; . \quad
\eea
Hereafter, the functionals \qref{d36} and \qref{d37} will be called
the single-particle and two-particle intracluster
functionals of CGF, respectively.

In analogy to \qref{d7}, from \qref{d20} and \qref{d23} one can obtain
expressions relating the intracluster $\Ff$-functions with their
functionals
\bea
\label{d38}
&&
\Ff_\nu\big( \{ \kt_\nu \}\big)  =
\Ff_\nu \big(\brepsn, \{ \kt_\nu \}\big)
\attwo{\kt^a_{\nu,\tau}=\kt^a_\nu }{\eps_\nu^a=0} \; ; \quad
\label{d39}
\\[7pt] &&
\Ff_{\nu r}\big(\{ {}^r \! \kt_\nu \brr
{}^\nu \! \kt_r \}\big)   =    \Ff_{\nu r}
\big(\brepsnr, \{ {}^r \mshalf \kt_\nu \brr
{}^\nu \! \kt_r \}\big)
\attwo{ {}^r \! \kt^a_{\nu,\tau}=
{}^r \mshalf \kt^a_\nu  \kaptwo }
{\eps_\nu^a \epszero = 0} \;\; . \quad
\eea
One can also derive expressions relating the single-particle and
two-particle  intracluster CGFs with their functionals
\bea
&& \hspace{-12mm}
\label{d40}
\acr{ \ttau \; \bar{S}_{\nu}^{a_1} (\tau_1) \;
\bar{S}_{\nu}^{a_2} (\tau_2) \;...\;  \bar{S}_{\nu}^{a_k} (\tau_k)
}{ \rho_\nu }=
\acr{ \ttau \; S_{\nu,\tau_1}^{a_1} \;
S_{\nu,\tau_2}^{a_2}  \;...\;  S_{\nu,\tau_k}^{a_k}
}{ \rho_\nu (\!\brepsn\!) } \;
\attwo{\kt^a_{\nu,\tau}=\kt^a_\nu }{\eps_\nu^a=0} \; , \quad
\\ && \hspace{-12mm}
\label{d41}
\acr{ \ttau \; \barbars_{\nu}^{a_1} (\tau_1) \;
\barbars_{\nu}^{a_2} (\tau_2) \;\ddd\;
\barbars_{\nu}^{a_k} (\tau_k) \;
\barbars_{r}^{a'_1} (\tau'_1) \;
\barbars_{r}^{a'_2} (\tau'_2)  \;\ddd\;
\barbars_{r}^{a'_l} (\tau'_l)   }{ \rho_{\nu r}}
\\[6pt] \nonumber &&  =
\acr{\ttau \; S^{a_1}_{\nu \tau_1} \; S^{a_2}_{\nu \tau_2} \;\ddd\;
S^{a_k}_{\nu \tau_k} \; S^{a'_1}_{r \tau'_1} \;
S^{a'_2}_{r \tau'_2} \;\ddd\; S^{a'_l}_{r \tau'_l}}
{\rho_{\nu r}(\!\brepsnr\!)} \;
\attwo{{}^r \mshalf \kt^a_{\nu,\tau}={}^r \mshalf \kt^a_\nu \kaptwo}
{\eps_\nu^a\epszero=0} \;\; ,
\eea
where
\bea
&& \hspace{-7mm}
\label{d42}
\bar{S}_\nu^{a} (\tau) =\ex{-\tau \ham_\nu } \;\msone
S_\nu^{a} \; \ex{\tau  \ham_\nu} \;; \qquad
\rho_\nu= \frac{ \ex{ \ham_\nu } }
{\Sp \big( \ex{ \ham_\nu } \big)  } \;; \qquad
\ham_\nu (\{ \kt_\nu \}) = \! \sum_a \kt_\nu^a S_\nu^a \;; \qquad
\\[7pt] && \hspace{-7mm}
\label{d44}
\bar{\!\bar{S}}_\nu^{a} (\tau) =
\ex{-\tau \ham_{\nu r} } \; S_\nu^{a} \;
\ex{\tau \ham_{\nu r}} \; ; \qquad
\rho_{\nu r}= \frac{ \ex{ \ham_{\nu r} } }
{\Sp \big( \ex{ \ham_{\nu r} } \big) } \;;
\\[7pt] && \nonumber \hspace{-7mm}
\ham_{\nu r} (\{ {}^r \mshalf \kt_\nu\brr {}^\nu \mshalf \kt_r\})
=\sum_a \big[ {}^r \mshalf \kt_\nu^a S_\nu^a +
{}^\nu \mshalf \kt_r^a S_r^a + \sum_b K^{ab} S_\nu^aS_r^b \big] \;.
\eea

From equations \qref{d27} - \qref{d29}, taking into account condition
of the extremum of the $\Fkor(\{\eps\})$-functional with respect to
$\fitn{r}{\nu,\tau}{a}$
\bea
\label{d46}
\hspace{2cm}
\prt{\Fkor\big(\{\eps\},\{\ae\brr\fii\}\big)}{\fitn{r}{\nu,\tau}{a}}=0
\; ,
\eea
we obtain the system of equations for the functionals
$\kar{\ttau S_{\nu,\tau}^a}{\krho(\!\{\eps\}\!)}$ and cluster
fields $\fitn{r}{\nu,\tau}{a}$.
\bea
\label{d47}
&&
\kar{\ttau S_{\nu,\tau}^a}{\krho(\!\{\eps\}\!)}  =
\Ff_\nu^{(1)}\big(\; {}^a_\tau  \big| \; \brepsn ,\{\kt_\nu\} \big)
\\[7pt]
\label{d48}
&&
\Ff_\nu^{(1)}\big(\; {}^a_\tau \big| \; \brepsn , \{\kt_\nu\} \big) =
\Ff_{\nu r}^{(1,0)} \big(\;{}^a_\tau \big|\; \brepsnr,
\{{}^r \msone \kt_\nu\brr {}^\nu \msone \kt_r\} \big) \;
\eea
One can see that equation \qref{d48} for
$\fitn{r}{\nu,\tau}{a}$, obtained from
the $\Fkor(\{\eps\})$-functional extremum condition
\qref{d46}, is equivalent to equations
\bea
\nonumber
\ang{\ttau S_{\nu,\tau}^a}_{\rho_\nu(\!\brepsn\!)}  =
\ang{\ttau S_{\nu,\tau}^a}_{\rho_{\nu r}(\!\brepsnr\!)} \; .
\eea
That is, in the present approximation (see \qref{d40} - \qref{d44})
the relations between the density matrices are not violated:
\[
\ang{S_{\nu}^a}_{\rho_\nu}  =
\ang{S_{\nu}^a}_{\rho_{\nu r}} \;\;\; \Longrightarrow \;\;\;
\rho_\nu = \Sp_{\sympm{S}_r} \!\; \rho_{\nu r} \;.
\]

Putting $\eps^a_\nu=0$ ($\ae^a_{\nu,\tau}=\ae^a_{\nu}$,
$\fitn{r}{\nu,\tau}{a}=\fitn{r}{\nu}{a}$,
see also \qref{d8}, \qref{d40}, \qref{d41}),
taking into account the following relations
\[
\ang{ S_{\nu}^a}_{\rho_\nu} =
\prt{\Ff_\nu(\{\kt_\nu\})}{\kt_\nu^a} \;, \qquad
\ang{ S_{\nu}^a}_{\rho_{\nu r}} =
\prt{\Ff_{\nu r}(\{{}^r \msone \kt_\nu\brr{}^\nu \msone \kt_r\})}
{{}^r \msone \kt_\nu^a} \;
\]
and going to the uniform fields case
$\ae^a_{\nu}=\ae^a$
($\fitn{r}{\nu}{a}=\fii^a$, $\kt_\nu^a=\kt^a$,
${}^{r} \kt_\nu^a=\ktt{}^a$), from \qref{d47}, \qref{d48}
we obtain the system of equations for the single-particle
distribution functions
$\kar{S^a}{\krho}$ and cluster fields $\fii^a$ in the TPCA
 for the uniform fields case.
\bea
\label{d47a}
&&
\kar{ S^a}{\krho} = \prt{\Ff_1(\{\kt\})}{\kt^a}
\\[7pt] &&
\label{d48a}
\prt{\Ff_1(\{\kt\})}{\kt^a}= \frac{1}{2}
\prt{\Ff_{12}(\{\ktt\})}{\ktt{}^a}
\eea
Here $\Ff_1(\{\kt\})$ and $\Ff_{12}(\{\ktt\})$ are the single-particle
and two-particle intracluster $\Ff$-functions for the uniform fields
case \qref{d20a}, \qref{d23a}. The factor
$\fr{1}{2}$ in the right hand side of equation \qref{d48a} arose at going
from the partial derivative of the
$\Ff_{\nu r}(\{{}^r\kt_\nu\brr{}^\nu\kt_r\})$-function with respect
to ${}^r\kt_\nu^a$ in the non-uniform fields case to the partial
derivative of the $\Ff_{12}(\{\ktt\})$-function with respect to
$\ktt{}^a$ in the uniform fields case.

\subsection{Pair distribution functions}

Let us briefly discuss the presented in  \cite{prhe}
method of calculation of the pair CGFs
functionals of the reference system, based on the technique
developed in \cite{pris} for the Ising model. Starting from
\qref{d13} and \qref{d47}, we obtain an expression for the pair CGF
functional.
\bea
\label{d51}
\!\!\!\!\!
\kora b^{(2)}
\big({}^{a_1}_{\nu,\tau_1} \bb {}^{a_2}_{\mu,\tau_2} \bb \{\eps\} \big)
\!=\!
\kacr{\ttau\;\! S^{a_1}_{\nu,\tau_1} \;\!
S^{a_2}_{\mu,\tau_2}}{\krho(\!\{\eps\}\!)}
\!=\!
\sum_{a_3} \! \int_0^1 \! {\rm d}\tau_3 \;
\Ff_\nu^{(2)} \big(\;\! {}^{a_1}_{\tau_1} \! \big| \;\;\!
{}^{a_3}_{\tau_3} \! \big|  \brepsn \big)   \!\cdot\!
\frac{ \delta \; \kt_{\nu,\tau_3}^{a_3}}
{ \delta \; \ae_{\mu,\tau_2}^{a_2} }
\eea

Having in mind calculations of the pair CGFs  (see \qref{d8})
for the uniform
\linebreak
f\/i\/e\/l\/d\/s~c\/a\/s\/e,
and since for specific
systems single-particle intracluster pair CGFs
$\acr{\ttau\;\bar{S}^{a_1}_{\nu}(\tau_1)\;
\bar{S}^{a_2}_{\nu}(\tau_2)}{\rho_\nu}$
(see \qref{d36}, \qref{d40}) in
uniform fields case can be calculated directly,
we need to obtain an equation for
${ \delta \; \kt_{\nu,\tau_3}^{a_3}}/
{ \delta \; \ae_{\mu,\tau_2}^{a_2} }$ \cite{pris,prhe,beg1}.

We introduce the notations
\bea
\label{d52}
\kt^\od_{\nu\mu}
\big({}^{a_3}_{\tau_3}\big|{}^{a_2}_{\tau_2}\big) =
\frac{ \delta \; \kt_{\nu,\tau_3}^{a_3}}
{ \delta \; \ae_{\mu,\tau_2}^{a_2} } \;\! ; \qquad
{}^r \msone \kt^\od_{\nu\mu}
\big({}^{a_3}_{\tau_3}\big|{}^{a_2}_{\tau_2}\big) =
\frac{ \delta \; {}^r \msone \kt_{\nu,\tau_3}^{a_3}}
{ \delta \; \ae_{\mu,\tau_2}^{a_2} } \;\! ; \qquad
\fitn{r}{\nu\mu}{\od}
\big({}^{a_3}_{\tau_3}\big|{}^{a_2}_{\tau_2}\big) =
\frac{ \delta \; \fitn{r}{\nu,\tau_3}{a_3}}
{ \delta \; \ae_{\mu,\tau_2}^{a_2} } \;\! . \quad
\eea
Taking the functional derivative
${ \delta \;  }/{ \delta \; \ae_{\mu,\tau_2}^{a_2} }$
from the both sides of equation \qref{d48}, and taking into
account the relation
\bea
\label{d53}
{}^r \msone \kt^\od_{\nu\mu}
\big({}^{a_3}_{\tau_3}\big|{}^{a_2}_{\tau_2}\big) =
\kt^\od_{\nu\mu}\big({}^{a_3}_{\tau_3}\big|{}^{a_2}_{\tau_2}\big) -
\fitn{r}{\nu\mu}{\od}
\big({}^{a_3}_{\tau_3}\big|{}^{a_2}_{\tau_2}\big)
\eea
(see \qref{d26}), we obtain
\bea
\label{d54}
&&
\sum_{a_3} \int_0^1 {\rm d} \tau_3 \;
\Ff_\nu^{(2)} \big( \;{}^{a_1}_{\tau_1} \! \big| \;\;
{}^{a_3}_{\tau_3} \! \big| \brepsn \big)   \cdot
\kt^\od_{\nu\mu}
\big({}^{a_3}_{\tau_3}\big|{}^{a_2}_{\tau_2}\big)
\\ && \nonumber \qquad
= \sum_{a_3} \int_0^1 {\rm d} \tau_3 \;
\Ff_{\nu r}^{(2,0)} \big( \;{}^{a_1}_{\tau_1} \! \big| \;\;
{}^{a_3}_{\tau_3} \! \big| \brepsnr \big)   \Big[
\kt^\od_{\nu\mu}
\big({}^{a_3}_{\tau_3}\big|{}^{a_2}_{\tau_2}\big) \! -
\fitn{r}{\nu\mu}{\od}
\big({}^{a_3}_{\tau_3}\big|{}^{a_2}_{\tau_2}\big) \Big]
\\ && \nonumber \qquad
+ \sum_{a_3} \int_0^1 {\rm d} \tau_3 \;
\Ff_{\nu r}^{(1,1)} \big( \;{}^{a_1}_{\tau_1} \! \big| \;\;
{}^{a_3}_{\tau_3} \! \big\| \brepsnr \big)   \Big[
\kt^\od_{r \mu}
\big({}^{a_3}_{\tau_3}\big|{}^{a_2}_{\tau_2}\big) \! -
\fitn{\nu}{r \mu}{\od} \big({}^{a_3}_{\tau_3}\big|{}^{a_2}_{\tau_2}\big)
\Big]  \; . \qquad
\eea
Going to a matrix form in \qref{d54}
%
%
\footnote{
Here the matrices have a block  structure; for instance in terms of
$(x,y,z)$:
\[
\wh{\Ff}_\nu^{(2)} \epsnbr =
\left(\begin{array}{lll}
\Ff_\nu^{(2)}(\;\!x|\;x|\brepsn\!)&\Ff_\nu^{(2)}(\;\!x|\;y|\brepsn\!)
&\Ff_\nu^{(2)}(\;\!x|\;z|\brepsn\!)\\
\Ff_\nu^{(2)}(\;\!y|\;x|\brepsn\!)&\Ff_\nu^{(2)}(\;\!y|\;y|\brepsn\!)
&\Ff_\nu^{(2)}(\;\!y|\;z|\brepsn\!)\\
\Ff_\nu^{(2)}(\;\!z|\;x|\brepsn\!)&\Ff_\nu^{(2)}(\;\!z|\;y|\brepsn\!)
&\Ff_\nu^{(2)}(\;\!z|\;z|\brepsn\!)
\end{array}\right) \; ,
\]
where the submatrices
\[
{\Ff}_\nu^{(2)}(\;\!a_1|\;a_2|\brepsn\!)\!=\!
\left(\! \begin{array}{lllll}
\!\Ff_\nu^{(2)}\big(\; {}^{a_1}_{0} \!\big|\;
{}^{a_2}_{0} \!\big|\brepsn\!\big)                      &\!
\!\Ff_\nu^{(2)}\big(\; {}^{a_1}_{0} \!\big|\;
{}^{a_2}_{{\rm d} \tau} \!\big|\brepsn\!\big)           &\!
\!\Ff_\nu^{(2)}\big(\; {}^{a_1}_{0} \!\big|\;
{}^{a_2}_{2{\rm d} \tau} \!\big|\brepsn\!\big)          &\!
\!\ddd                                                 &\!
\!\Ff_\nu^{(2)}\big(\; {}^{a_1}_{0} \!\big|\;
{}^{a_2}_{1} \!\big|\brepsn\!\big)                      \\
\vdots &&&&                                            \\
\!\Ff_\nu^{(2)}\big(\; {}^{a_1}_{1} \!\big|\;
{}^{a_2}_{0} \!\big|\brepsn\!\big)                      &\!
\!\Ff_\nu^{(2)}\big(\; {}^{a_1}_{1} \!\big|\;
{}^{a_2}_{{\rm d} \tau} \!\big|\brepsn\!\big)           &\!
\!\Ff_\nu^{(2)}\big(\; {}^{a_1}_{1} \!\big|\;
{}^{a_2}_{2{\rm d} \tau} \!\big|\brepsn\!\big)          &\!
\!\ddd                                                 &\!
\!\Ff_\nu^{(2)}\big(\; {}^{a_1}_{1} \!\big|\;
{}^{a_2}_{1} \!\big|\brepsn\!\big)
\end{array}\!\!\right)
\]
are the ${\rm M}\times{\rm M}$ matrices
(${\rm M}=\frac1{{\rm d}\tau}+1$).
At ${\rm d}\tau \longrightarrow 0$ ${\rm M} \longrightarrow \infty$
}
%
%
and performing some transformations, we obtain:
\bea
\label{d55} &&
\Big[ \wh {\Ff}_{\nu r}^{(2,0)} \epsnrbr
- \wh {\Ff}_{\nu}^{(2)} \epsnbr \Big] \cdot
\wh{\kt}{}^\od_{\nu \mu} +
\wh {\Ff}_{\nu r}^{(1,1)} \epsnrbr  \cdot
\wh{\kt}{}^\od_{r \mu}
\\[7pt] \nonumber &&
= \wh {\Ff}_{\nu r}^{(2,0)} \epsnrbr \cdot
{}^r \msone \wh {\fii}^\od_{\nu \mu}
+ \wh {\Ff}_{\nu r}^{(1,1)} \epsnrbr \cdot
{}^\nu \msone \wh {\fii}^\od_{r \mu} \;\; .
\eea
Introducing the notations
\bea
\label{d56} &&
\wh {f}_{\nu r} = \big[
\wh {\Ff}_{\nu r}^{(2,0)} \epsnrbr \big]^{-1} \cdot
\wh {\Ff}_{\nu r}^{(1,1)} \epsnrbr \;\! ;
\\[6pt] \nonumber &&
\wh {v}_{\nu r} = \big[
\wh {\Ff}_{\nu r}^{(2,0)} \epsnrbr \big]^{-1} \cdot
\wh {\Ff}_{\nu}^{(2)} \epsnrbr \; ,
\eea
we rewrite equation \qref{d55} as
\bea
\label{d57}
{}^r \msone \wh {\fii}^\od_{\nu \mu} +
\wh {f}_{\nu r} \cdot {}^\nu \msone \wh {\fii}^\od_{r \mu} =
\wh {f}_{\nu r} \cdot \wh{\kt}{}^\od_{r \mu}
+\big(\wh{1} - \wh{v}_{\nu r} \big) \wh{\kt}{}^\od_{\nu \mu}  \; ,
\qquad (r \in \pi_\nu).
\eea
We obtain an equation \qref{d57} with unknown
${}^r \msone {\fii}^\od_{\nu \mu}$,
${}^\nu \msone {\fii}^\od_{r \mu} $.
One more linearly independent equation still should be derived.
After changing indices
$r \rightleftharpoons \nu$, we get
\bea
\label{d58}
{}^\nu \msone \wh {\fii}^\od_{r \mu} +
\wh {f}_{r \nu} \cdot {}^r \msone \wh {\fii}^\od_{\nu \mu} =
\wh {f}_{r \nu} \cdot \wh{\kt}{}^\od_{\nu \mu}
+\big(\wh{1} - \wh{v}_{r \nu} \big) \wh{\kt}{}^\od_{r \mu}  \; ,
\qquad (\nu \in \pi_r).
\eea
One can easily see that \qref{d57} and \qref{d58} are a system of
equations for  ${}^r \msone {\fii}^\od_{\nu\mu}$,
${}^\nu \msone {\fii}^\od_{r \mu}$.
Summing up over $r \in \pi_\nu$  in \qref{d57} and taking into
account the fact that
\bea
\label{d59}
\wh{\kt}{}^\od_{\nu\mu} =
\delta_{\nu\mu} \cdot \wh{1} +
\sum_{r \in {\textstyle \pi_\nu}}
{}^r \msone \wh {\fii}^\od_{\nu\mu} \; ,
\eea
from the system of equations \qref{d57}, \qref{d58} one obtains
a closed equation for ${\kt}{}^\od_{\nu\mu}$.
\bea
\label{d60} &&
\Big\{ \wh{1} + \! \sum_{r \in {\textstyle \pi_\nu}} \!
\wh{f}_{\nu r}
\Big[\wh{1}-\wh{f}_{r \nu}\!\cdot\! \wh{f}_{\nu r}\Big]^{\!-1}
\Big[ \wh{f}_{r \nu} -
\wh{f}_{\nu r}^{-1} \big(\;\! \wh{1} - \wh{v}_{\nu r} \big) \Big]
\! \Big\}
\; \wh{\kt}{}^\od_{\nu\mu}
\\ \nonumber && \qquad \quad
=\delta_{\nu\mu} \!\cdot\! \wh{1} +
\sum_{r \in {\textstyle \pi_\nu} }
\wh{f}_{\nu r}
\Big[\wh{1}-\wh{f}_{r \nu}\!\cdot\! \wh{f}_{\nu r}\Big]^{\!-1}
\;\! \wh{v}_{r \nu} \cdot \wh{\kt}{}^\od_{r \mu} \quad
\eea

Let us put $\eps^a_\nu=0$ and go to the uniform fields case
$\ae^a_{\nu}=\ae^a$
\bea
\label{d62}
&& \hspace{-5mm}
\wh{\Ff}_{\nu r}^{(2,0)} \epsnrbr
= \wh{\Ff}^{(2,0)}_{(\nu-r)} \epsnrbr
\longrightarrow \wh{\Ff}^{(2,0)}_{12} \;\! ; \qquad
\wh{\Ff}_{\nu}^{(2)} \epsnbr
\longrightarrow \wh{\Ff}_1^{(2)} \;\! ; \qquad
\\[5pt] && \nonumber \hspace{-5mm}
\wh{\Ff}_{\nu r}^{(1,1)} \epsnrbr
= \wh{\Ff}^{(1,1)}_{(\nu-r)} \epsnrbr
\longrightarrow \wh{\Ff}^{(1,1)}_{12} \;\! ; \qquad
\wh{f}_{\nu r}= \wh{f}_{(\nu-r)} \longrightarrow \wh{\fmi}  \;\! ;
\\[5pt] && \nonumber \hspace{-5mm}
\wh{v}_{\nu r}= \wh{v}_{(\nu-r)} \longrightarrow \wh{v} \; \qquad
\eea
in equation \qref{d60}.
Then it can be rewritten as
\bea
\label{d63}
\Big[\;\!\wh{1}+(z\!-\!1)
\wh{\fmi}^{\;2} -z\big(\!\;\wh{1}-\wh{v}\big)\Big]
\; \wh{\kt}{}^\od_{\nu\mu} =
[\;\! \wh{1} - \wh{\fmi}^{\;2}] \;
\delta_{\nu\mu}+\wh{\fmi} \cdot\wh{v} \;\!
\sum_{r=1}^N \pi_{\nu r}\cdot\wh{\kt}{}^\od_{r \mu} \; ,
\eea
where
\bea
\label{d64}
\pi_{\nu r} = \left\{ \begin{array}{l} 1, \;\;\;\;
 r \in  \pi_\nu \\
0, \;\;\;\; r \notin \pi_\nu \end{array} \right. \; .
\eea
It should be remembered, that
with putting $\eps^a_\nu=0$
we go from the single-particle and two-particle
intracluster CGF functionals (see \qref{d36}, \qref{d37})
to the corresponding CGFs (see \qref{d40}, \qref{d41}).

Going to the frequency-momentum representation in \qref{d63} and
solving the obtained equation, we get for
$\wh{\kt}{}^\od(\vq ,\omn)$
\bea
\label{d65} \hspace{-10mm}
\wh{\kt}{}^\od(\vq, \omn) &=& \Big[ \wh{1}+
(z\!-\!1) \wh{\fmi}^{\;2}(\omn)-z \big[\;\!\wh{1}-\wh{v}(\omn) \big] -
\wh{\fmi}(\omn) \!\cdot\! \wh{v}(\omn) \!\cdot\!\pi( \vq )
\Big]^{-1}
\\ \nonumber \hspace{-10mm}
&\times& \big[\;\!\wh{1} - \wh{\fmi}^{\;2}(\omn) \big] \;.
\eea
Here
\bea
\label{d65a}
\wh{\fmi} (\omn)= [\fgfm{12}{2,0}{\omn} ]^{-1}
\cdot \fgfm{12}{1,1}{\omn} \;;
\qquad
\wh{v} (\omn)= [\fgfm{12}{2,0}{\omn} ]^{-1}
\cdot \fgfm{1}{2}{\omn} \; \quad
\eea
and $\fgfm{12}{2,0}{\omn}$, $\fgfm{12}{1,1}{\omn}$, $\fgfm{1}{2}{\omn}$
are $3\times3$ matrices in the indices $a,b$ ($a=x,y,z$ or $+,-,z$),
their elements are Fourier transforms
($\fgfdoom{a}{b}$, $\fgfdrom{a}{b}$, $\fgfoom{a}{b}$)
of the pair intracluster CGFs
$\sdc{\ttau \; \barbars_1^a(\tau) \; \barbars_1^b(0)}$,
$\sdc{\ttau \; \barbars_1^a(\tau) \; \barbars_2^b(0)}$,
$\soc{\ttau \bar{S}_1^a(\tau) \bar{S}_1^b(0)}$, respectively.
$\pi(\vq )$ is the Fourier transform of the function $\pi_{\nu r}$.
For simple lattices with a hypercubic symmetry, $\pi(\vq )$ reads
\bea
\label{d66}
\pi(\vq )=2\sum_{i=1}^d \cos(q_i \cdot \alpha) \;;
\eea
$d$ is the lattice dimensionality;
$\alpha$ is the lattice constant.
The obtained matrix expression \qref{d65} can be rewritten as
\bea
\label{d67} && \hspace{-15mm}
\wh{\kt}{}^\od(\vq ,\omn) \msone=\! \Big[
z \wh{v}(\omn) \msone-\msone
(z\!-\!1) \big[\;\!\wh{1}\msone+\msone \wh{\fmi}(\omn) \big] +
z\big[\;\!\wh{1} \msone-\msone \wh{\fmi}(\omn) \big]^{-1} \;\!
\wh{\fmi}(\omn) \!\cdot\! \wh{v}(\omn) \!\cdot\! \Theta(\vq )
\msone \Big]^{\msone -1} \;\;\;
\\ \nonumber && \; \times \;\;\!
\big[\;\!\wh{1}+\wh{\fmi}(\omn)\big] \;\!,
\eea
where $\Theta(\vq )$ for simple lattices with a hypercubic symmetry is
\bea
\label{d69}
\Theta(\vq ) =1-\frac{\pi(\vq )}{z}=\frac{2}{d} \sum_{i=1}^d
\sin^2 \left( \frac{q_i \cdot \alpha}{2} \right)  \; .
\eea

Putting $\eps^a_\nu=0$, going to the uniform fields case in relation
\qref{d51}, and going to the frequency-momentum representation, we
obtain expressions for pair CGFs, which are convenient to rewrite
in a matrix form in the indices $a$, $b$:
\bea
\label{d70}
\korb \wh{b}^{(2)}(\qomn)=\wh{\Ff}_1^{(2)}(\omn) \cdot
\wh{\kt}{}^\od(\qomn) \;.
\eea

Hence, in order to calculate the pair CGFs of the reference system in
the uniform fields case from \qref{d70}, we need to
calculate (according to \qref{d67} and \qref{d65a}) the single-
and two-particle intracluster pair CGFs
$\fgfoom{a}{b} \msone=\msone%
\socom{\ttau \bar{S}_1^a(\tau) \bar{S}_1^b(0)}$,
$\fgfdoom{a}{b}=\sdcom{\ttau \; \barbars_1^a(\tau) \; \barbars_1^b(0)}$,
$\;\fgfdrom{a}{b}=\sdcom{\ttau \; \barbars_1^a(\tau) \; \barbars_2^b(0)}$.

\section{Ising model in transverse field}
\setcounter{equation}{0}

\subsection{Thermodynamics. General results}
We consider the Ising model in transverse field with
a renormalized pseudospin operator
($S^z=(-1, 1)$).
\bea
\label{3'1}
H = - \sum_{\nu=1}^N (h S_\nu^z + \Gamma  S_\nu^x)
- \fr{1}{2} \sum_{\nu,\delta} K S_\nu^z S_{\nu+\delta}^z
- \fr{1}{2}\sum_{\nu,\mu} J_{\nu\mu} S_\nu^z S_\mu^z
\eea
Here $K$ and $J_{\nu\mu}$ are the short-range and long-range pair
interactions, respectively; $\Gamma$ is the transverse field;
the quantity $h\rightarrow0$ is introduced for the sake of
convenience. Hereafter, the factor $\beta=(k_BT)^{-1}$ is
written explicitly.

In the framework of MFA for the long-range interactions, the Hamiltonian
\qref{3'1} can be written as
\bea
\label{3'2}
H = \kora H + \fr{1}{2} N J_0 m^2 \; ,
\eea
where
\bea
\label{3'2a}
J_0=\sum_{\mu=1}^N J_{\nu\mu} \;, \qquad m=\langle S^z \rangle_{\!\rho}
\eea
and $\kora H$ is the Hamiltonian of the reference IMTF
\bea
\label{3'3}
&&
\kora H =
-\sum_{\nu=1}^N \Big[ \ae^z S_\nu^z + \ae^x S_\nu^x \Big]
-\fr{1}{2}
\sum_{\nu,\delta} K S_\nu^z S_{\nu+\delta}^z
 \; ;
\\[6pt]
\label{3'4}
&&
\ae^z = h + J_0 m \; ; \qquad   \ae^x = \Gamma  \; .
\eea
According to the results of previous sections, the free
energy of IMTF within the TPCA for the short-range interactions, with the
long-range interactions taken into account within the MFA, is
\bea
\label{3'17}
F = -k_B T \cdot \Fkor + \fr{1}{2} N J_{0} m^2 .
\eea
The $\Fkor$-function of the reference IMTF
\bea
\label{3'19}
\Fkor =(1-z) N \Ff_1 + \fr{zN}{2} \Ff_{12}
\eea
is expressed via the single-particle
\bea
\label{3'20}
&&
\Ff_1 = \ln Z_1 \;, \qquad  Z_1=
{\rm Sp}_{ \sympm{S}_1} \ex{-\beta H_1} \; ,
\\[6pt]
\label{3'21}
&&
H_1 = -
\sum_{a=x,z} \kta S_1^a \;\; ; \qquad
\kt^a = \ae^a + z \fii^a \;\; , \qquad
(a=z,x) \;\; ,
\eea
and two-particle
\bea
\label{3'22}
&& \hspace{-16mm}
\Ff_{12} = \ln Z_{12} \;, \qquad Z_{12}=
{\rm Sp}_{ \sympm{S}_1, \sympm{S}_2}
{\rm e}^{-\beta H_{12}} \; ; \;\;\;\;\;\;
\\[6pt]  && \hspace{-16mm}
\label{3'23}
H_{12} = - \! \sum_{a=x,z}
\ktta \big( S_1^a + S_2^a \big) - K S_1^z  S_2^z \;\! ;
\qquad\!\!
\ktt{}^a =\ae^a +(z\!-\!1) \fii^a \!\; , \quad (a=z,x) \;
\eea
intracluster $\Ff$-functions. Let us show briefly how these functions
can be obtained.

The Hamiltonian $H_1$ acts on the basis of two functions of state of a
single particle
\bea
\label{3'24}
\begin{array}{cc} 1 & + \\ 2 & - \end{array}
\eea
In the representation \qref{3'24}, the single-particle Hamiltonian reads
\bea
\label{3'25}
H_1=  -
\left( \begin{array}{cc} \ktz & \ktx \\ \ktx & -\ktz \end{array}
\right) .
\eea
Taking into account \qref{3'20}, one can easily obtain the
single-particle partition function in an explicit form
\bea
\label{3'26}
Z_1 = 2 {\rm ch} (\beta \Lambda)  \;\; ; \qquad
\Lambda=\sqrt{ (\ktz)^2 +(\ktx)^2 } \; .
\eea

The two particle Hamiltonian $H_{12}$ acts on the basis of four
functions of state of a two-particle cluster.
\bea
\label{3'27}
\begin{array}{ccc}
1 & + & + \\ 2 & + & - \\ 3 & - & + \\ 4 & - & -
\end{array}
\eea
In the representation \qref{3'27}, the Hamiltonian $H_{12}$ reads
\bea
\label{3'28}
H_{12}= - \left(
\begin{array}{cccc}
2 \kttz  +  K & {} \kttx & {} \kttx & {} 0
\vspace{0.25cm} \\
\kttx & {} - K & {} 0 & {} \kttx
\vspace{0.25cm} \\
\kttx & {} 0 & {} - K & {} \kttx
\vspace{0.25cm} \\
0 & {} \kttx & {} \kttx & {} - 2 \kttz  + K
\end{array}
\right) \;.
\eea
On the basis of \qref{3'22} and \qref{3'28} we obtain the two-particle
partition function
\bea
\label{3'29}
Z_{12} =
\sum_{i=1}^4 {\rm e}^{- \beta \etpn{i}}   \; ,
\eea
where
\bea
\label{3'30}
{\textstyle \etp{\;\!4} = K } \; ,
\eea
whereas three other eigenvalues  $\etp{\;\!1}$,
$\etp{\;\!2}$, $\etp{\;\!3}$
of the matrix \qref{3'28} are roots of a cubic equation
\bea
\label{3'31} \!\!\!\!\!\!
E_{12}^3 + K E_{12}^2 - \!
\Big[ K^2+4 \big(\kttx \big)^2
\msone+ 4 \big(\kttz \big)^2 \Big] E_{12}
-K \Big[ K^2 \msone+
4 \big(\kttx \big)^2 \msone-4 \big(\kttz \big)^2 \Big]
\msone=\msone 0 \;\! .
\eea

From \qref{d47a} and \qref{d48a}, with taking into account the fact
that  in the framework of the MFA for the long-range interactions
\bea
\label{3'34a}
\ang{S^z}_{\!\rho}=-\frac{1}{N}\poch{F}{h}=
\korang{S^z}_{\krho}=-\frac{1}{N}\poch{\kora F}{\ae^z}
\eea
(this can be obtained from the explicit expression for the free energy
\qref{3'17}), we get equations
for the parameters $m=\ang{S^z}_{\!\rho}$,
$\etax=\ang{S^x}_{\!\rho}$ and cluster fields
$ \fii^{a}$ $(a=z,x)$.
\bea
&& \hspace{-14mm}
\label{3'35}
\frac{\ktx}{\Lambda} \th(\beta \Lambda) = \frac{4\kttx}{Z_{12}}
\sum_{i=1}^3 \frac{ [-\etpn{\ii} - K] {\rm e}^{-\beta \etpn{i}} }
{3 \etpn{\ii}^2 + 2K \etpn{\ii} - [ K^2 + 4(\kttx)^2 + 4(\kttz)^2]}
\\ && \hspace{-14mm}
\label{3'36}
\frac{\ktz}{\Lambda} \th(\beta \Lambda) = \frac{4\kttz}{Z_{12}}
\sum_{i=1}^3 \frac{ [-\etpn{\ii} + K] {\rm e}^{-\beta \etpn{i}} }
{3 \etpn{\ii}^2 + 2K \etpn{\ii} - [ K^2 + 4(\kttx)^2 + 4(\kttz)^2]}
\\ && \hspace{-14mm}
\label{3'37}
m=\frac{\ktz}{\Lambda} \th(\beta \Lambda)
\\ && \hspace{-14mm}
\label{3'38}
\etax=\frac{\ktx}{\Lambda} \th(\beta \Lambda)
\eea
When the long-range interaction is absent ($J_0=0$),
we have a system of two equations \qref{3'35} and \qref{3'36} for
$ \fix  $, $ \fiz  $ in an implicit form
($\etpn{\ii}$ are roots of cubic equation
\qref{3'31}) and expressions \qref{3'37} for $m$ and \qref{3'38} for
$\etax$.
When  $J_0 \neq 0$, we have a system of three equations \qref{3'35} --
\qref{3'37} for $ \fix $, $ \fiz $, and $m$, and an expression
for $\etax$.


Numerical analysis of the thermodynamic
characteristics and  longitudinal static
susceptibility $\chi^{zz}$
(which is too cumbersome to be presented here)   obtained here within
the TPCA in the short-range interactions, with the long-range interactions
taken into account within the MFA, as well as study of the
applicability bounds
of these approach to the IMTF on different types of lattices
at different values of the parameters
$\Gamma$, $J_0$ will be given elsewhere. Here we shall only briefly
consider the major results at
$K\msone>\msone0$, $J_0\msone \geq \msone0$, $\Gamma\msone>\msone0$.
We shall use the terminology of ferroelectricity.

For the one-dimensional IMTF at
$J_0\msone=\msone0$, the two-particle cluster
approximation, unlike the MFA for the short-range interactions, does not
predict existence of ferroelectric ordering (at
$T\msone>\msone0$ and
arbitrary $\Gamma$ the system is in the paraelectric phase).
Comparison of the TPCA results for the free energy, entropy, and specific
heat as functions of temperature
(expressions for entropy and specific heat were obtained for the
paraelectric phase only) at different values of $\Gamma/K$
has shown, that this approximation yields fair results for these
characteristics at all temperatures except for the low-temperature region.
Thus, at high temperatures, the TPCA results accord with exact
results not only qualitatively, but also well enough quantitatively.
The lower temperature, the more results of TPCA differ from exact ones
(too lower values of free energy, entropy, and specific heat), whereas
in the low-temperature region
$T\msone<\msone{T_l}$
($k_BT_l/K\msone<\msone%
\th(\sqrt[4]{\fr{1}{6}\!\cdot\msone
\Gamma/K})+\fr{1}{6}\!\cdot\msone \Gamma/K$)
are qualitatively incorrect (for instance,
the free energy is an increasing function of temperature).

For the one-dimensional model at
$J_0\msone>\msone0$, as well as for
two-dimensional and three-dimensional models at
$J_0\msone\geq\msone0$,
the TPCA for the short-range interactions with,
the long-range interactions
taken into account within the MFA, predicts that
a limiting  value $(\Gamma/K)_k$ exists,
which depends on $J_0$ and $z$, and
above which a ferroelectric ordering is impossible (the latter is a
qualitatively correct result). At
$(\Gamma/K)_a\msone<\msone\Gamma/K\msone<\msone(\Gamma/K)_k$
(where $(\Gamma/K)_a\msone=%
\msone\sqrt{c(z,\Gamma,J_0,K)\!\cdot\!zJ_0/K}$,
$c(z,\Gamma,J_0,K)\msone\approx\msone2$)
this approximation predicts a phase transition from the paraelectric
phase to the ferroelectric phase on lowering temperature and the phase
transition from the ferroelectric to the paraelectric phase -- the
so-called anti-Curie point. At small enough values of
$\Gamma/K\msone\leq\msone(\Gamma/K)_a$
the anti-Curie point is absent, but the
temperature behavior of the thermodynamic characteristics remains
qualitatively incorrect. The low-temperature region of
$T\msone<\msone{T_l}$, where the
TPCA yields incorrect results for thermodynamic characteristics,
is reduced
when the  values of $\Gamma/K$ and $J_0/K$ decrease (see Table~1.).
\begin{table}[htb]
\caption{Temperature of the anti-Curie point $T_a$, temperature $T_l$
below which unphysical results for $m(T)$ and $\chi^{zz}(T)$
are obtained, and Curie temperature for a square lattice ($z=4$)
at different values of $\Gamma$ and $J_0$ within the TPCA for the
short-range interactions, calculated with the long-range
interactions taken into account in the MFA.
\protect\newline $~$ }
\centerline{
\begin{tabular}{|l|l|l|l|l|}
\hline
$k_BT_a/K$ &$k_BT_l/K$ &$k_BT_c/K$ &$J_0/K$ &$\Gamma/K$ \\
\hline
0.01 & 0.57   & 2.86   & 0.0   & 0.5    \\
0.26 & 0.93   & 2.40   & 0.0   & 2.0    \\
0.03 & 0.96   & 3.04   & 0.4   & 2.0     \\
\hline
\end{tabular} }
\end{table}
In the high-temperature region
$T\msone>\msone{T_l}$ at $J_0\msone=\msone0$ and
$z\msone>\msone2$, the TPCA
is much more correct than the mean field approximation for the
short-range interactions.

We also performed a numerical analysis of the TPCA results at neglecting
the variational parameter $\fii^x$ ($\fii^x\!=\!0$). This version of the
approximation is not suitable for one-dimensional chains. Thus, at small
enough values of  $J_0/K\msone<\msone0.09$ and
$\Gamma/K$, Curie temperature increases on increasing $\Gamma/K$. At
$J_0=0$ and $\Gamma/K~\!\in~]0,\;\!1.28]$
a ferroelectric ordering is predicted.

For two-dimensional and three-dimensional lattices, neglecting the
variational parameter $\fii^x$ leads to a slight quantitative
worsening of the results in a high-temperature region and to  a
qualitatively correct description of temperature dependences of
thermodynamic characteristics ($m(T)$, $\chi^{zz}(T)$) in a
low-temperature region.
This worsening is the smaller, the larger are the lattice
dimensionality and the value of the long-range interaction,
and the smaller is the transverse field.

\subsection{Thermodynamics and intracluster pair
distribution functions  in paraelectric phase}
In order to study the dynamic characterictics of the IMTF in the
paraelectric phase, we write here certain relations for some
thermodynamic quantities in the paraelectric phase
($\ae^z=\ktz=\kttz=\fiz=0$, $m=0$).
The eigenvalues $\etpn{\ii}$ of two-particle Hamiltonian \qref{3'28}
in the paraelectric phase  are (see \qref{3'31}):
\bea
\label{3'46}
\etp{\;\!1}= - \tildL \; ; \qquad
\etp{\;\!2}= \tildL \; ; \qquad
\etp{\;\!3}= K \; ; \qquad  \etp{\;\!4}= - K \; ,
\eea
where
\bea
\label{3'47}
\tildL = \sqrt{K^2+4{(\kttx)}^2} \; .
\eea
From \qref{3'46} we obtain the two-particle partition function in the
paraelectric phase explicitly
\bea
Z_{12}=
2\Big[ \ch(\beta \tildL ) + \ch(\beta K) \Big].
\eea

In the paraelectric phase, equations \qref{3'36}, \qref{3'37} turn to
identity, while equation \qref{3'35} for the variational parameter $\fix$
can be written explicitly, using \qref{3'46}
\bea
\label{3'49}
\th(\beta \ktx)= \frac{4 \kttx}{\tildL Z_{12}}
\sh(\beta \tildL ) \; .
\eea
We also present here an expression for
$\etax=\ang{S^x}_{\msone\rho}$:
\bea
\label{3'50}
\etax=\th(\beta \ktx)  \;.
\eea

To calculate the pair CGFs of the reference model \qref{3'3} within
the TPCA (see \qref{d70}), we need to know the single-particle and
two-particle intracluster pair CGFs
$\fgfoom{a}{b}$, $\fgfdoom{a}{b}$, $\fgfdrom{a}{b}$ ($a,b=x,y,z$).
Let us calculate now the two-particle  incracluster CGFs. It is convenient
to do so in the self-representation of the operator $H_{12}$.

Since we have explicit expressions for the eigenvalues of the two-particle
Hamiltonian \qref{3'28} in the paraelectric phase (see \qref{3'46}),
it is easy to obtain the normalized unitary matrix, which diagonalizes the
two-particle Hamiltonian
\bea
\label{3'54}
{\hat U}= \left(  \begin{array}{cccc}
r_1 & r_2 & 0 & {1}/{\sqrt2} \\ r_2 & -r_1 & {1}/{\sqrt2} & 0 \\
r_2 & -r_1 & -{1}/{\sqrt2} & 0 \\ r_1 & r_2 & 0 & -{1}/{\sqrt2}
\end{array} \right) \; .
\eea
Here we use the notations
\bea
\label{3'55}
r_1=\fr12 \cdot \sqrt{1+{K}/{\tildL}} \;; \qquad
r_2=\fr12 \cdot \sqrt{1-{K}/{\tildL}} \;.  \qquad
\eea

Going from the Pauli operators to their four-row analogs
\cite{he32,he33,he34}
\bea
\label{3'52}
\si_1^a=S_1^a \otimes I \; ; \qquad
\si_2^a=I \otimes S_2^a \; ; \qquad a=x,y,z
\eea
(here $I$ is the two-row unit matrix, $\otimes$ is the direct
product symbol; matrices $\si_\nu^a$ obey Pauli
commutation rules) and performing a unitary transformation
\bea
\label{3'53}
\sit_\nu^a={\hat U}^{-1} \si_\nu^a \psone {\hat U},
\eea
we obtain the pseudospin operators in a self-representation of the
operator $H_{12}$:
\bea
\label{3'56}
&& \hspace{-12mm}
\sit_1^z=\sqrt2 \left( \begin{array}{cccc}
0 & 0 & r_2 & r_1   \\   0 & 0 & -r_1 & r_2    \\
r_2 & -r_1 & 0 & 0    \\  r_1 & r_2 & 0 & 0
\end{array} \right) \;\! ; \qquad
\sit_2^z=\sqrt2 \left( \begin{array}{cccc}
0 & 0 & -r_2 & r_1   \\   0 & 0 & r_1 & r_2    \\
-r_2 & r_1 & 0 & 0    \\  r_1 & r_2 & 0 & 0
\end{array} \right) \!\; ; \\ && \nonumber \hspace{-12mm}
\sit_1^x= \left( \! \begin{array}{cccc}
{2\kttx}/{\tildL} & - {K}/{\tildL} & 0 & 0   \\
-{K}/{\tildL} & -{2\kttx}/{\tildL} & 0 & 0    \\
0 & 0 & 0 & -1    \\    0 & 0 & -1 & 0
\end{array} \! \right) \;\! ; \qquad\!\!\!\!
\sit_2^x= \left( \! \begin{array}{cccc}
{2\kttx}/{\tildL} & -{K}/{\tildL} & 0 & 0   \\
-{K}/{\tildL} & -{2\kttx}/{\tildL} & 0 & 0    \\
0 & 0 & 0 & 1    \\    0 & 0 & 1 & 0
\end{array} \! \right) \;\! ; \\ && \nonumber \hspace{-12mm}
\sit_1^y=i \sqrt2 \left( \begin{array}{cccc}
0 & 0 & r_1 & r_2   \\   0 & 0 & r_2 & -r_1    \\
-r_1 & -r_2 & 0 & 0    \\  -r_2 & r_1 & 0 & 0
\end{array} \right) \;\! ; \qquad
\sit_2^y=i \sqrt2 \left( \begin{array}{cccc}
0 & 0 & -r_1 & r_2   \\   0 & 0 & -r_2 & -r_1    \\
r_1 & r_2 & 0 & 0    \\  -r_2 & r_1 & 0 & 0
\end{array} \right) \;\! .
\eea
Expanding operators $\sit_i^a$ \qref{3'56} in finite series in the
four-dimensional Hubbard operators \cite{he32,he33,he34},
following \cite{pr4,pr58}, we easily calculate the
two-particle cumulant pair intracluster Green functions:
\bea
\label{3'57}
&&
\fgfm{12}{2,0}{\omn}= \left( \begin{array}{ccc}
\fgfdoom{x}{x}&0&0 \\ 0&\fgfdoom{y}{y}&\fgfdoom{y}{z} \\
0&\fgfdoom{z}{y}&\fgfdoom{z}{z}
\end{array} \right) \; ,
\\ && \nonumber
\fgfm{12}{1,1}{\omn}= \left( \begin{array}{ccc}
\fgfdrom{x}{x}&0&0 \\ 0&\fgfdrom{y}{y}&\fgfdrom{y}{z} \\
0&\fgfdrom{z}{y}&\fgfdrom{z}{z}
\end{array} \right) \; ,
\eea
where
\bea
\label{3'58}
&& \hspace{-13mm}
\fgfdoom{x}{x}=\sdcom{\ttau \sit_1^x(\tau) \sit_1^x(0)}=
A_s \delta(\omn)+A_{\scriptscriptstyle +}
(\omn) \; \; \\ && \nonumber \hspace{-13mm}
\fgfdoom{y}{y}=\frac4{\beta \tildL Z_{12} \psi(\omn)}
\big[ {(2\kttx)}^4 \sh(\beta \tildL ) +
C_{\scriptscriptstyle \!+} \cdot \omn^2 \big] \; ;
\\ && \nonumber \hspace{-13mm}
\fgfdoom{y}{z}=-\fgfdoom{z}{y}=
\frac{8\kttx \omn}{\beta \tildL Z_{12} \psi(\omn)}
\big[ C_{\scriptscriptstyle \!-}
+ \sh(\beta \tildL ) \omn^2 \big] \; ;
\\ && \nonumber \hspace{-13mm}
\fgfdoom{z}{z}=\frac{16{(\kttx)}^2}{\beta \tildL Z_{12} \psi(\omn)}
\big[C_{\scriptscriptstyle \!-}
+ \sh(\beta \tildL )  \omn^2 \big] \; ;
\\[9pt] && \hspace{-13mm}
\label{3'59}
\fgfdrom{x}{x}=\sdcom{\ttau \sit_1^x(\tau) \sit_2^x(0)}=
A_s \delta(\omn)+A_{\scriptscriptstyle -}(\omn)
\; \; \\ && \nonumber \hspace{-13mm}
\fgfdrom{y}{y}=-\frac{16BK\omn^2}{\beta Z_{12} \psi(\omn)} \; ;
\\ && \nonumber \hspace{-13mm}
\fgfdrom{y}{z}=-\fgfdrom{z}{y}=
\frac{32BK\kttx\omn}{\beta Z_{12} \psi(\omn)} \; ;
\\ && \nonumber \hspace{-13mm}
\fgfdrom{z}{z}=
\frac{64BK{(\kttx)}^2}{\beta Z_{12} \psi(\omn)} \; .
\eea
Here we use the notations
\bea
\label{3'60}
&&
A_s=\Big(\frac{4\kttx}{\tildL Z_{12}}\Big)^{\!\!2} \cdot
\big[1+\ch(\beta K) \ch(\beta \tildL ) \big] \; ;  \qquad
B=\fr12 [\ch(\beta \tildL ) - \ch(\beta K)] \; ; \qquad
\\[7pt] && \nonumber
A_{\scriptscriptstyle \pm}(\omn)=\frac{8K}{\beta \tildL Z_{12}} \Big(
\frac{K \sh(\beta \tildL )}{4 \tildL ^2 +\omn^2} \pm
\frac{\tildL \sh(\beta K)}{4 K^2 +\omn^2}
\Big) \; ;
\\[7pt] && \nonumber
C_{\scriptscriptstyle \!\pm}=[\tildL ^2 +K^2] \sh(\beta \tildL ) \pm
2 \tildL K \sh(\beta K) \; ;
\label{3'61}
\\[7pt] &&
\psi(\omn) = \big[ (\tildL +K)^2 + \omn^2 \big]
\big[ (\tildL -K)^2 + \omn^2 \big] \; .
\eea

The most convenient way to obtain the single-particle
intracluster CGFs is, by performing a rotation in a spin space,
to go to such a coordinate system, where Hamiltonian  \qref{3'25}
is diagonal. We present here the final result
(after the inverse transformation) in the paraelectric phase
in terms of $a=x,y,z$:
\bea
\label{3'63}
\fgfm{1}{2}{\omn}= \left( \begin{array}{ccc}
\big[ 1-\etax^2 \big] \delta(\omn) &0&0 \\
0& g(\omn) & g'(\omn) \\ 0& -g'(\omn) & g(\omn)
\end{array} \right) .
\eea
Here we use notations
\bea
\label{3'64}
g(\omn)= \frac{4}{\beta} \cdot
\frac{\etax \cdot \ktx}{{(2\ktx)}^2+\omn^2} \; ; \qquad
g'(\omn)= \frac{\omn}{2\ktx} \cdot g(\omn) \; .
\eea
At obtaining \qref{3'63} we used the fact that within TPCA for the
short-range interactions, with taking into account the long-range
interactions in the MFA, $\so{S^x}=\ang{S^x}_{\!\rho}\equiv\etax$.

\subsection{Dynamics in paraelectric phase.
Cluster random phase approximation}

Our task is to investigate the dynamics characteristics
of the IMTF in the paraelectric phase
within the TPCA for the short-range interactions and within
the $(r_0^{-d})^0$ approximation for the long-range interactions
\cite{prhe,prdg} -- the cluster random phase approximation.
The first step is then to calcualte the temperature CGFs.

In CRPA the pair CGF $\fgm(\qomn)\equiv {\wh b}^{(2)}(\qomn)$
according to \qref{2.31} is
\bea
\label{3'65}
\fgm(\qomn)=\big[1-\fgmk(\qomn) \beta \hat{J}( \vq) \big]^{-1} \psone
\fgmk(\qomn) \; .
\eea
where
\bea
\label{3'66}
\hat{J}( \vq)= \left( \begin{array}{ccc}
0& 0 &0 \\ 0 &0&0\\ 0&0& J( \vq) \end{array} \right) \; ,
\eea
(in terms of $a=x,y,z$), and
$\fgmk(\qomn)\equiv\kora {\wh b}^{(2)}(\qomn)$ is the pair CGF of the
reference system, which in the TPCA reads \qref{d70}.

From \qref{3'65}, \qref{d70}, using
\qref{3'57}, \qref{3'63},
we obtain pair CGFs. For $\fg^{xx}(\qomn)$,
and $\fg^{zz}(\qomn)$ in the paraelectric phase we have:
\bea
\label{3'71}
\fg^{xx}( \qomn) =\delta(\omn) \cdot \fg^{xx}_\alpha( \vq ) \;, \qquad
\fg^{zz}( \qomn)=\frac{4\Gamma\etax
[p_{\!_+}+\omn^2] [p_{\!_-}+\omn^2]}
{R( \qomn)}.
\eea
Here we introduce the notations
\bea
\label{3'72}
&& \hspace{-13mm}
\fg^{xx}_\alpha( \vq )=
\big[1\!-\!\etax^2\big] \left\{
\frac{z\big[1\!-\!\etax^2\big]}{d_x (T)} -(z-1) + z
\big[1\!-\!\etax^2\big]
\frac{b_x(T)}{d_x(T)} \Theta( \vq) \right\}^{-1} \;;
\\ \nonumber && \hspace{-13mm}
d_x(T)=\frac{K^2}{\tildL ^2} \cdot \frac{\etax}{\beta \kttx}
+2 \Bigl[1-\frac{K^2}{\tildL ^2}\Bigr]
\frac{1+\ch( \beta\tildL ) \ch (\beta K)}
{\ch (\beta \tildL ) +\ch (\beta K)} \; ;
\\ && \nonumber \hspace{-13mm}
b_x(T)=\frac{1}{2\sh(\beta K)} \left\{ \frac{K^3}{\tildL ^3}
\sh(\beta\tildL ) - \sh (\beta K) + \beta K \Big[1-
\frac{K^2}{\tildL ^2}\Big]
\frac{1+\ch( \beta K) \ch (\beta \tildL )}
{\ch (\beta\tildL ) + \ch (\beta K)}\!\right\} ;
\\[7pt] && \hspace{-13mm}
\label{3'74}
R( \qomn )= [p_{\!_-}+\omn^2]
\Big[ \omn^4 + {\rm u}_2 \;\! \omn^2 +
{\rm u}_0  - 4 \Gamma \etax J( \vq )
[p_{\!_+} + \omn^2] \Big]
\\ \nonumber && \qquad
+\frac{4 B\tildL \Gamma K}{\kttx \sh(\beta\tildL )} \!\cdot\!
\psi (\omn) z\Theta( \vq ) ;
\\[7pt] && \hspace{-13mm}
\nonumber
p_{\!_{\pm}} \msone=\msone K^2 \!+\! \tildL ^2 \msone
+ \frac{2 \tildL K \big[\!-\sh (\beta K) \pm 2B \big]}
{\sh(\beta\tildL )}  \;\!; \quad
p_{s} \msone=\msone K^2 \!+\! \tildL ^2 \msone
+ \frac{2 \tildL K \big[\sh (\beta K) \!-\! 2B \big] }
{\sh(\beta\tildL )}  \;\!;
\\ && \hspace{-13mm} \nonumber
{\rm u}_2 \!=\! 2z^2 [K^2 \!+\! \tildL ^2] \msone+\msone (z\!-\!1)^2
\big[(2\ktx)^2 \!+\msone p_{\!_+}  \big]-
2z(z\!-\!1) \Big\{ \msone \frac{\ktx
\big[ (2\kttx)^2 \!+ \msone p_{s} \big]}{2\kttx}
\msone +  p_{\!_+} \msone \Big\} ;
\\ && \nonumber \hspace{-13mm}
{\rm u}_0 =4\Gamma \{ 4z ( \kttx)^3 -(z\!-\!1)\ktx p_{\!_+} \} \;.
\eea
In calculations of \qref{3'71}
we used the relations \qref{3'49}, \qref{3'50}.

It should be noted that from \qref{3'71} one obtains the static
longitudinal susceptibility ($\chi^{zz}=\beta\fg^{zz}(0,0)$) of the IMTF
in the paraelectric phase
\bea
\label{3'80}
\chi^{zz} = \left[\frac{z\tildL Z_{12}}
{2 \{ \frac{\tildL +K}{\tildL -K} \cdot \ex{\beta \tildL }-
\frac{\tildL -K}{\tildL +K} \cdot \ex{-\beta \tildL }-
\frac{4\tildL K}{{\tildL }^2-K^2} \cdot \ex{\beta K} \}  }
-\frac{(z\!-\!1)\ktx}{\th (\beta \ktx)} - J_0 \right]^{-1} \;,
\eea
($J_0=J(\vq \!=\!0)$)
which accords with the one calculated from thermodynamic relations in the
TPCA for the short-range interactions, with the long-range interactions
taken into account in the MFA.

To explore the dynamic properties of the IMTF we need to know
not the temperature CGFs \qref{3'71}, but the retarded CGFs. We can
calculate them \cite{Izum} by performing
analytical continuation
of the temperature CGFs $\fg^{ab}(\qomn)$
($i\omn\rightarrow E+iE'$) and going to
the limit $E'\rightarrow 0$. The final
results for spectral densities
${\cal J}^{xx}( \vq, E)$ and ${\cal J}^{zz}( \vq, E)$,
defined as
\bea
\label{3'81}
{\cal J}^{ab}( \vq, E) = \lim_{E'\to0} \left[
\frac{2 \hbar \beta}{\ex{\beta E}-1}
{\rm Im} \psone \fg^{ab}( \vq , \omn)
_{\big|_{\scriptstyle{\omn \rightarrow -iE+E'}}}
\right] \;,
\eea
and for the pair cumulant correlation function
$\ang{S_{\vq }^z S_{-\vq }^z}{}^{\! c}$ are the following.

The spectral density ${\cal J}^{xx}( \vq, E)$ of the IMTF in the
paraelectric phase within the CRPA reads
\bea
\label{3'81a}
{\cal J}^{xx}( \vq, E) =
\delta(E) \cdot \fg^{xx}_\alpha( \vq ) \;.
\eea
Let us note, that an exact expression
for ${\cal J}^{xx}( \vq, E)$ of the one-dimensional
IMTF with the short-range interactions only \cite{pr63}
has not only the central peak ($\sim \delta(E)$) but also two
symmetrical resonance zones. Absence of the resonance
zones within the CRPA for ${\cal J}^{xx}( \vq, E)$
results from neglecting fluctuations of cluster fields in this
approximation.

The spectral density ${\cal J}^{zz}( \vq, E)$ (in the paraelectric phase)
can be presented as
\bea
\label{3'83}
\fr1{\hbar} {\cal J}^{zz}( \vq, E) = \sum_{i=-,+,r} k^J_i( \vq)
\Big[ \delta(E-E_i( \vq )) + \ex{\beta E_i( \vq )}
\delta(E+E_i( \vq )) \Big] \; ,
\eea
where $E_i( \vq )$ ($i=-,+,r$) are the elementary excitation spectrum
modes, determined from the equation
(see \qref{3'71}, \qref{3'74})
\be
\label{3'77}
R( \vq , \omn)
_{\big|_{\scriptstyle{\omn \rightarrow -iE}}} =0 \;,
\ee
and $k^J_i( \vq )$ are the integral intensities of the
elementary excitations spectrum modes
\bea
\label{3'84}
k^J_i( \vq)= 4 \pi T \;\! \Gamma \;\! \etax \;
{\cal A}_i( \vq ) \!\cdot\!
\frac{[p_{\!_+}-E_i^2( \vq )][p_{\!_-} -E_i^2( \vq )] }
{ (\ex{\beta E_i( \vq )}-1) E_i( \vq )} \;.
\eea
Here we use the notations:
\bea
\label{3'85}
{\cal A}_{i_1}( \vq )
= \frac{1}{[E_{i_1}^2( \vq ) -E_{i_2}^2( \vq )]
[E_{i_1}^2( \vq ) -E_{i_3}^2( \vq )] } \; , \qquad
\begin{array}{l}
i_1,i_2,i_3 = (-,+,r) \; , \\
i_1 \neq i_2, \;\; i_2 \neq i_3, \;\; i_3 \neq i_1 \; .
\end{array}
\eea
It should be noted that at $\vq =0$
(see \qref{3'71}, \qref{3'74}) there are only two modes
\bea
E_{\pm} (0) = \frac{1}{\sqrt 2}\sqrt{ {\rm U}_2 \pm
\sqrt{{\rm U}_2^2-4 {\rm U}_0} } \;\! ,
\eea
where
\be
\label{3'79}
{\rm U}_2 ={\rm u}_2 -4 \etax \Gamma J_0; \qquad
{\rm U}_0 ={\rm u}_0 - 4\etax \Gamma J_0 \;\! p_{\!_+} .
\ee
The mode $E_-( \vq )$ is soft ($E_-(0) \rightarrow 0$, $T
\rightarrow T_c$).

For the pair cumulant correlation function
$\ang{S_{\vq }^z S_{-\vq }^z}{}^{\! c}$ from
\bea
\label{3'86}
\ang{S_{\vq }^a S_{-\vq }^b}{}^{\! c} {}
_{\big|_{\scriptstyle{t \longrightarrow 0}}}
=\fr1\hbar \int_{-\infty}^\infty {\rm d} E
\;{\cal J}^{ab}( \vq,E) \; ,
\eea
we obtain
\bea
\label{3'87}
\ang{S_{\vq }^z S_{-\vq }^z}{}^{\! c} {}
_{\big|_{\scriptstyle{t \longrightarrow 0}}}
=
2 T \;\! \Gamma \;\! \etax \sum_{i=-,+,r}
{\cal A}_i( \vq ) \!\cdot\! {\rm cth}(\fr12 \beta E_i)  \!\cdot\!
\frac{[p_{\!_+}-E_i^2( \vq )][p_{\!_-} -E_i^2( \vq )] }{E_i( \vq )} \; .
\eea


\renewcommand{\vq}{q}

\begin{figure}[htb]
\begin{center}
\leavevmode
\mframe{\epsffile{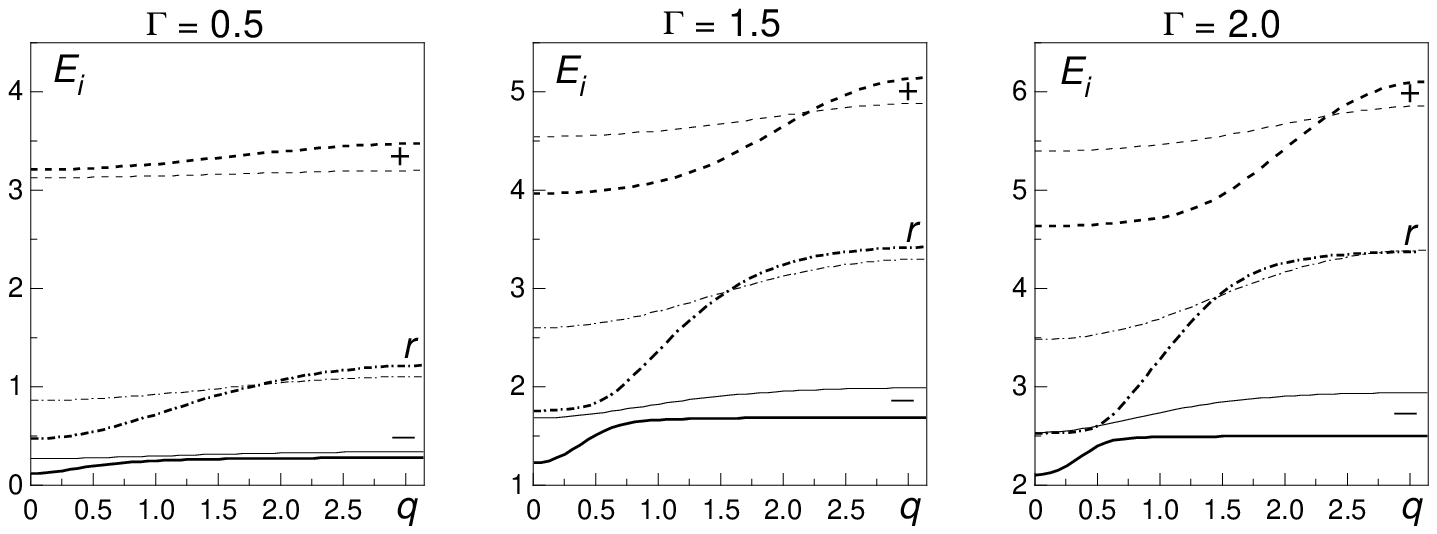}} %
\\ \vspace*{5mm}
\mframe{\epsffile{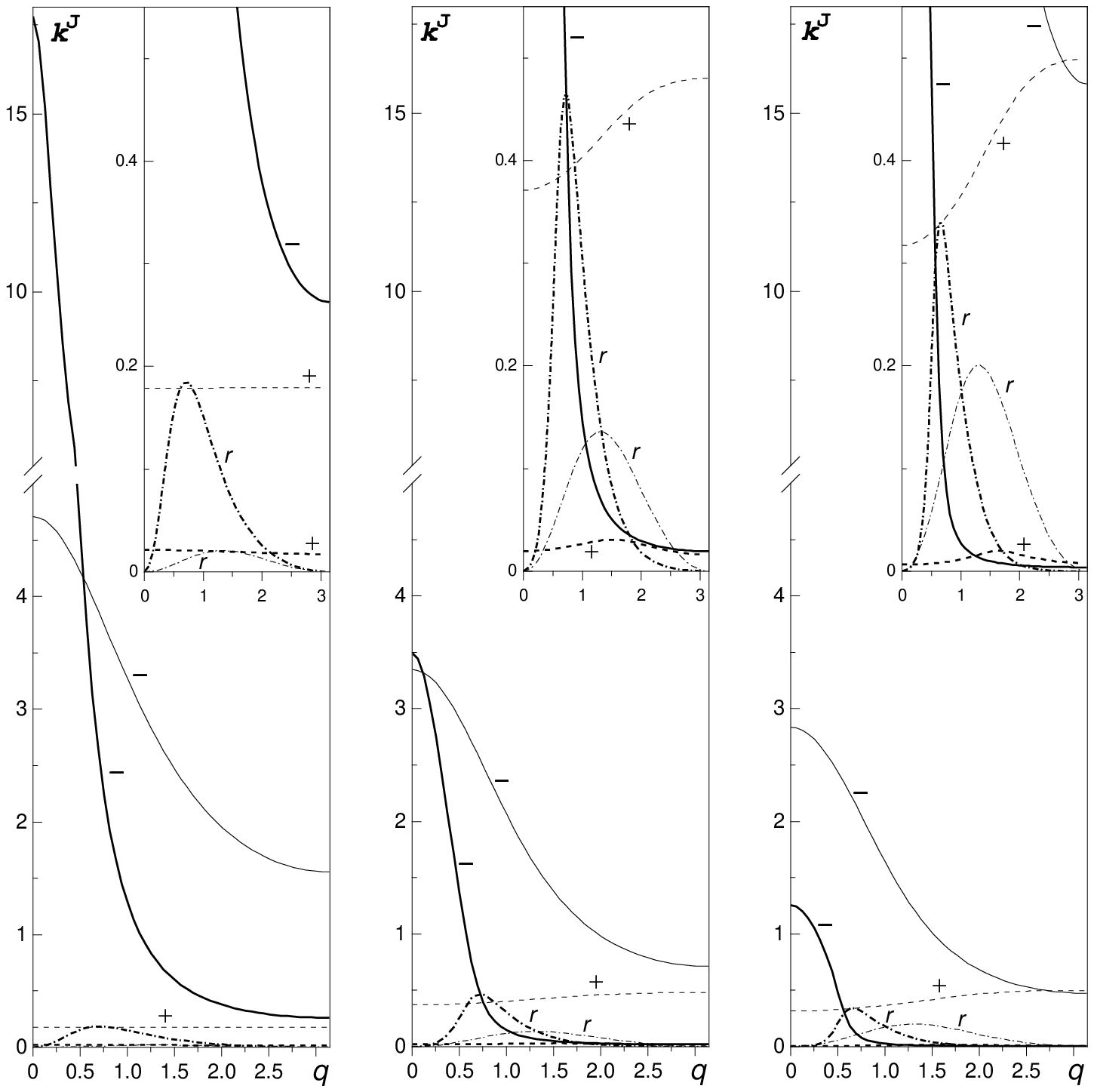}} %
\end{center} \vspace{1mm}
\unitlength=1cm
\vspace{-5mm}
\caption{ Elementary excitations spectrum modes $E_{i}( \vq )$
and their integral intensities $k^J_{i}( \vq )$ (within CRPA)
as functions of quasimomentum $\vq$ at different temperatures
(thick lines -- $k_BT=1.0$, thin lines -- $k_BT=4.0$) for
$\Gamma=0.5,\; 1.5, \;2.0$.}
\label{fig1}
\end{figure}

Let us briefly consider results of numerical analysis of the longitudinal
characteristics of the IMTF at $z=2$, $J_{\nu \mu}=0$ ($K=1$).
As we have already
mentioned, within CRPA the spectrum of the longitudinal characteristics
of the model  contains three nondamping modes $E_-( \vq)$, $E_+(
\vq)$, $E_r( \vq)$ with the integral intensities $k^J_-(\vq)$,
$k^J_+(\vq)$ and $k^J_r(\vq)$ in the spectral density
${\cal J}^{zz}( \vq,E)$. The calculated dependences of $E_i( \vq)$,
$k^J_i(\vq)$ at different transverse fields and temperatures are
presented in fig. \ref{fig1}.
On increasing $\Gamma$, temperature, and $\vq$, redistribution
of the intensities from low frequencies to higher frequencies
is observed. At large $\Gamma$ and low temperatures
the redistribution on increasing $\vq$ takes place,
first, mainly from $E_-( \vq)$ to
$E_r( \vq)$, and then from $E_r( \vq)$ to $E_+( \vq)$.
At large  $\Gamma$ and high temperatures the redistribution
takes place mainly from $E_-( \vq)$ to $E_+( \vq)$.
At small $\Gamma$ the redistribution is practically absent.

\begin{figure}[htb]
\hspace*{20mm} $\langle S_\nu^z S_{\nu+n}^z \rangle$
\hspace*{45mm} $\langle S_\nu^z S_{\nu+n}^z \rangle$
\vspace{-2mm}
\begin{center}
\leavevmode
\mframe{\epsffile{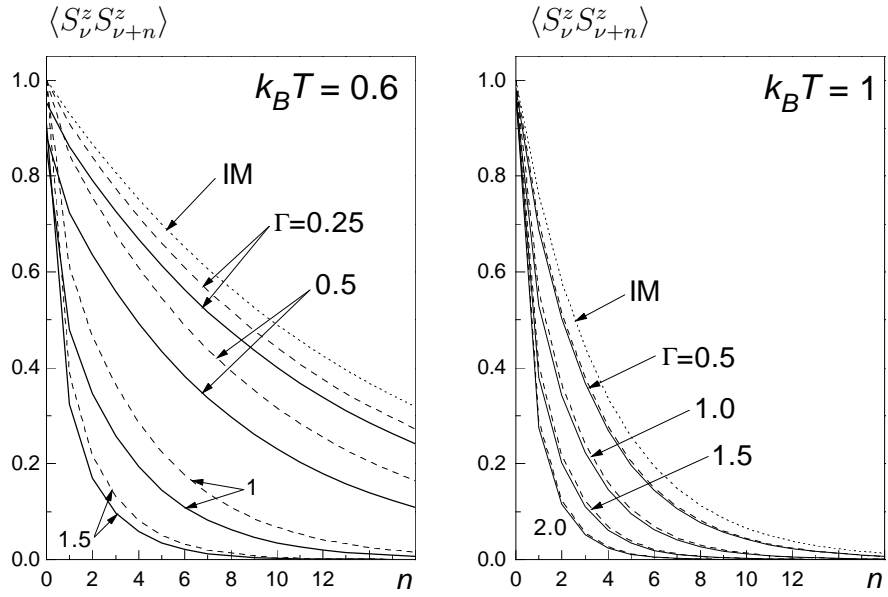}} %
\end{center} \vspace{1mm}
\unitlength=1cm
\vspace{-1mm}
\caption{Static correlation function
$\langle S^z_\nu S^z_{\nu+n} \rangle$
at different values of $\Gamma$
and  temperature ($k_BT=0.6,\;1.0$) calculated
within CRPA (solid lines)
and exactly \protect{\cite{Alla}} (dash lines).
Exact and approximate (CRPA) results for Ising model (short
dash lines) coincide. } \label{fig2} \end{figure}

In fig. \ref{fig2}  we present exact and approximate (CRPA) results for
the static correlator $\ang{S^z_\nu S^z_{\nu+n}}$
at $k_BT=0.6$ and $k_BT=1.0$
at different  $\Gamma$.
The CRPA gives too low values of
$\ang{S^z_\nu S^z_{\nu+n}}$, especially at low temperatures. Thus, the
autocorrelator $\ang{S^z_\nu S^z_\nu}$ at low temperatures is essentially
less than unity. The higher temperature and smaller $\Gamma$, the better
the CRPA results accord with exact ones.

\begin{figure}[htb]
\hspace*{2mm}
$2 \Gamma \!\cdot\! \RekPsifull, \; k^J_i(0)$
\vspace{-2mm}
\begin{center}
\leavevmode
\mframe{\epsffile{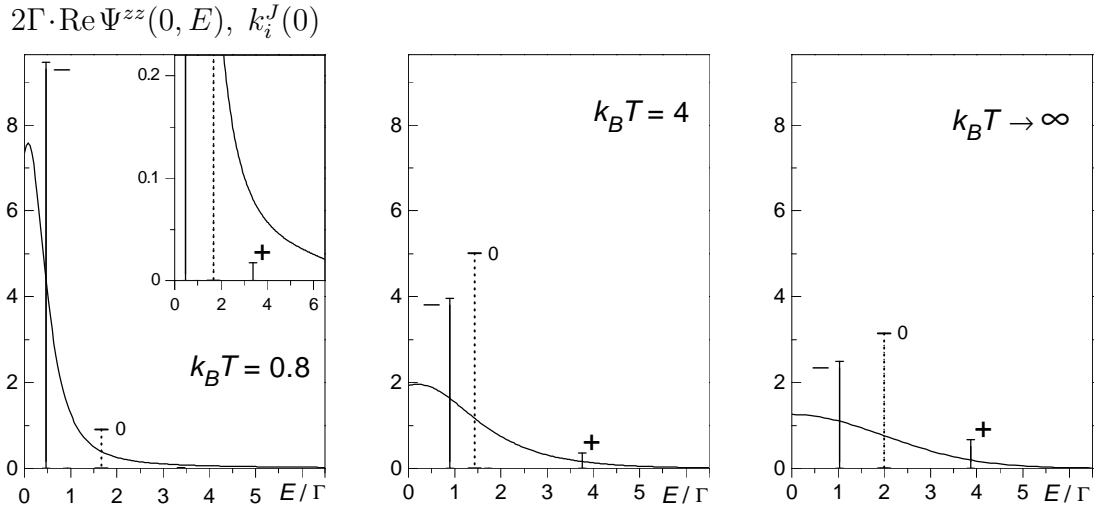}} %
\end{center} \vspace{1mm}
\unitlength=1cm
\vspace{-1mm}
\caption{Frequency dependence of
$2 \Gamma \!\cdot\! \RekPsifull$ for $\Gamma=1.0$ at different
temperatures ($k_BT=0.8,\;4,\;\infty$) calculated numerically
\protect{\cite{Oleg,r39}}. Vertical lines correspond to the mode
integral intensities $k^J_-(0)$, $k^J_+(0)$ within CRPA (solid line)
and $k^J_0(0)$ within RPA (dash line).}
\label{fig3}
\end{figure}

\begin{figure}[htb]
\hspace*{2mm}
$2 \Gamma \!\cdot\! \RekPsifull, \; k^J_i(0)$
\vspace{-2mm}
\begin{center}
\leavevmode
\mframe{\epsffile{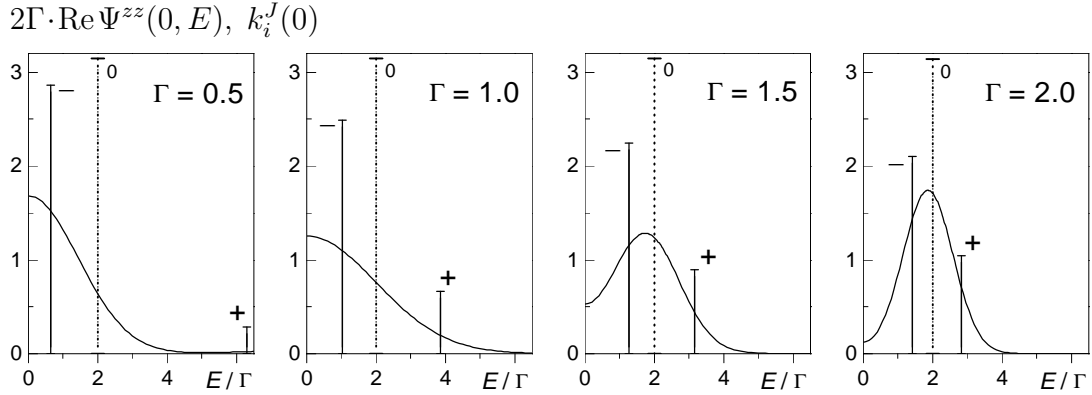}} %
\end{center} \vspace{1mm}
\unitlength=1cm
\vspace{-1mm}
\caption{Exact results for frequency dependence of
$2 \Gamma \!\cdot\! \RekPsifull$ at 
$T \rightarrow \infty$ at
different values of $\Gamma$ ($\Gamma=0.5,\;1.0,\;1.5,\;2.0$).
Vertical lines correspond to the mode integral
intensities $k^J_-(0)$, $k^J_+(0)$ within CRPA (solid line)
and $k^J_0(0)$ within RPA (dash line).}
\label{fig4}
\end{figure}

Let us discuss also the redistribution of the modes intensities
$E_-(0)$, $E_+(0)$ on changing $\Gamma$ and temperature.
At temperatures $k_B T > \sqrt{\Gamma}$,
change of the modes positions and integral
intensities (obtained within the CRPA) on changing $\Gamma$ and $T$
qualitatively describes change in the frequency dependences of the
real part of the relaxation function $\RekPsifull$, calculated
numerically \cite{Oleg,r39} or exactly.
Thus, for instance, at $\Gamma=1$ and $k_BT=0.8$
(see fig. \ref{fig3})
$\RekPsifull$ has a prominent resonance zone
at $E$ close to zero.
Smearing  of this resonance zone on increasing $T$ is qualitatively
described by increasing $k^J_+(0)$ and
$E_-(0)$ and by decreasing $k^J_-(0)$.
On the other hand, for instance, at $T\rightarrow\infty$
(see fig. \ref{fig4}), a shift of the resonance zone
$\RekPsifull$ to higher frequencies region on increasing
$\Gamma$ is described by increasing
\linebreak

\clearpage

\vspace{-5mm}

\vspace{-5mm}

\clearpage

\noindent
$E_-(0)$ and $k^J_+(0)$.
The fact that the resonance zone at
$\Gamma=0.5$ is more prominent than at $\Gamma=1.0$ is
described by decreasing $k^J_-(0)$ and
increasing $k^J_+(0)$ on increasing $\Gamma$.
The fact that at $\Gamma=2.0$
the resonance zone $\RekPsifull$   is more prominent than at $\Gamma=1.5$
is described  by  approaching of the modes positions
$E_-(0)$, $E_+(0)$ on increasing $\Gamma$.
In figs. \ref{fig3}, \ref{fig4} we depicted also the results of
the random phase approximation
(RPA) for the short-range interactions.  This approximation describes the
frequency dependences of the real part of the relaxation function in the
presented in figs. \ref{fig3}, \ref{fig4} cases qualitatively
well only at $T\to\infty$ $\Gamma=1.5,\; 2.0$.

We restricted the presented here  numerical analysis of the longitudinal
characteristics of the one-dimensional IMTF (at $J_{\nu \mu}=0$)
by the values of the transverse field $0.2\leq \Gamma\leq 2$.  A more
detailed analysis both at $J_{\nu \mu}=0$ and at $J_{\nu \mu}\ne0$ and
with the values of the microparameters corresponding to the CsH$_2$PO$_4$
crystal will be performed elsewhere.

\end{document}